\documentclass[pre,nofootinbib,superscriptaddress,twocolumn,floatfix]{revtex4}
\usepackage{graphicx}
\usepackage{amsmath,units}
\usepackage{mathrsfs}
\usepackage{color} 
\usepackage{cancel}
\usepackage{titlesec}
\usepackage[export]{adjustbox}
\usepackage{tikz}
\usepackage[section]{placeins} %added
\allowdisplaybreaks[1]

\newcommand{\rt}[1]{\textcolor{black}{#1}}
\begin{document}

\title{Emergent clustering due to \rt{quorum sensing} interactions in active matter}

\author{Samudrajit Thapa}
\affiliation{School of Mechanical Engineering, Tel Aviv University, Tel Aviv 69978, Israel}
\affiliation{Center for Computational Molecular and Materials Science, Tel Aviv University, Tel Aviv 69978, Israel}
\affiliation{Max Planck Institute for the Physics of Complex Systems, Nöthnitzer Straße 38, 01187 Dresden, Germany}
\author{Bat-El Pinchasik}
\affiliation{School of Mechanical Engineering, Tel Aviv University, Tel Aviv 69978, Israel}
\affiliation{Center for the Physics and Chemistry of Living Systems, Tel Aviv University, 69978, Tel Aviv, Israel}
\author{Yair Shokef}
\email{shokef@tau.ac.il}
\affiliation{School of Mechanical Engineering, Tel Aviv University, Tel Aviv 69978, Israel}
\affiliation{Center for Computational Molecular and Materials Science, Tel Aviv University, Tel Aviv 69978, Israel}
\affiliation{Center for the Physics and Chemistry of Living Systems, Tel Aviv University, 69978, Tel Aviv, Israel}
\affiliation{International Institute for Sustainability with Knotted Chiral Meta Matter, Hiroshima University, Japan}

\date{\today}
%\date{January 17, 2024}

\begin{abstract}
Many organisms in nature use local interactions to generate global cooperative phenomena. To unravel how the behavior of individuals generates effective interactions within a group, we introduce a simple model, wherein each agent senses the presence of others nearby and changes its physical motion accordingly. This generates non-physical, or virtual interactions between agents. We study the radial distribution function and the cluster size distribution to quantify the emergent interactions for both social and anti-social behavior; We identify social behavior as when an agent exhibits a tendency to remain in the vicinity of other agents, whereas anti-social behavior as when it displays a tendency to escape from the vicinity of others. Using Langevin dynamics simulations in two and three spatial dimensions, we discover that under certain conditions, positive correlations, which indicate attraction can emerge even in the case of anti-social behavior. Our results are potentially useful for designing robotic swimmers that can swim collectively only based on sensing the distance to their neighbors, without measuring any orientational information. 
\end{abstract}

\maketitle

\section{Introduction}

Interactions between group members have become a central field of study within the subject of active matter. The reason for that lies in the benefit that large groups have over individuals in nature. For example, in order to avoid being eaten in the case of locusts swarms and fish schools~\cite{partridge1982, ariel2015}, improving feeding efficiency in the case of bird flocks~\cite{sridhar2009} and ants colonies~\cite{bonabeau2000}, and even finding a new nesting place in the case of social bees~\cite{lindauer1961}. Social behavior is also an indication to the animal's well-being and normal function~\cite{brodin2013}. \rt{Additionally, numerous bacteria possess the ability to biochemically  perceive the local density of either the same or different species in their surroundings via \textit{quorum sensing}~\cite{bassler1999bacteria,bassler2002small,bassler2006bacterially,waters2005quorum,atkinson2009quorum,mukherjee2019bacterial,diggle2007quorum,brown2001cooperation,crespi2001evolution}. This mechanism has been widely studied in the context of regulating cooperative phenomenon such as bio-film formation~\cite{parsek2005sociomicrobiology,kindler2019burst}. Quorum sensing has also been studied in the context of synaptic signalling~\cite{ram2018smaller,cohen2010quorum}. Decision making processes resulting in cooperative behavior based on quorum sensing have been investigated also for higher organisms such as ants~\cite{pratt2002quorum,franks2015ants} and bees~\cite{seeley2004quorum,seeley2006group}.}

These impressive demonstrations of cooperative motion have led researchers to mimic such natural systems, wishing to add group ``intelligence'' to synthetic systems including small particles and robots. However, groups of engineered members vary in number, size of each member, propulsion, communication schemes and sensing capabilities. The size of each group member in artificial groups ranges between micrometer scale particles and swimmers~\cite{jin2021} to centimeter scale robots~\cite{rubenstein2012}. Nevertheless, similar physical laws can be applied to describe the dynamics of large groups~\cite{helbing1995, zuriguel2014}. Specifically, in groups of actively moving members one can apply statistical mechanics in order to understand and predict the group motion and density~\cite{bialek2012}. These works are related to the ongoing efforts to understand how information based interactions affect physical behavior of smart active systems~\cite{BenYaakov, Golkov, smart-active-matter, informational-active-matter}.

Physics-inspired Vicsek-like models of active Brownian particles (ABP) have been widely employed to investigate emergent collective phenomena -- such as clustering or flocking -- in different systems of self propelled particles~\cite{vicsek1995, romanczuk2012, vicsek2012, lemaitre2023, volpe}. Similarly, models of active Ornstein-Uhlenbeck particles have also been investigated~\cite{caprini2019active,caprini2022dynamics}. While many of these mathematical models are based on social interactions in the form of local velocity alignment of individuals~\cite{vicsek2012}, even without alignment, individuals can form clusters due to \textit{motility induced phase separation} (MIPS)~\cite{cates2008,cates2015motility, marchetti2012, merrigan2020,iyer2023dynamics,farrell2012pattern,paoluzzi2020information}. Clustering without alignment has also been observed for ABPs interacting repulsively~\cite{buttinoni2013}. Furthermore, alternative active-matter models which incorporate position-based interactions -- instead of velocity alignment -- have shown a broad range of emergent collective phenomena~\cite{huepe2014, peruani2016}.

Cooperative phenomena emerge in many living systems due to perception-based changes in the dynamics at the individual level~\cite{couzin2002, kay2008, pearce2014, gorbonos2016}. Specifically, switching between an active (non-zero velocity) and a passive state (null velocity) depending on feedback from visual perception within an interaction range (often a visual cone) can result in group formation~\cite{bechinger2019,zhou2023quorum}. It has also been shown that in the absence of alignment, selective social (attractive) and anti-social (repulsive) behavior depending on visual perception results in swarming and pattern formation~\cite{romanczuk2012}. However, this requires the particles to be able  to distinguish between approaching and escaping neighbors within the perception range. Collective motion arising from purely attractive or repulsive attractions have also been studied~\cite{romey1996, strmbom2011, vicsek2012}. 

In this article, we present a minimal model of ABP without orientational alignment and we investigate simple, short-ranged social and anti-social interaction rules that give rise to emergent clustering. We consider the behavior of a particle to be social when it exhibits a tendency to move more slowly when in the vicinity of other particles, whereas anti-social behavior is when it moves more rapidly and thus has a tendency to escape from the vicinity of other particles. In addition to confirming that positive correlations result from social behavior while negative correlations arise from anti-social behavior, we show that clustering can emerge even in case of anti-social behavior and even with just a few particles.  

\section{Active Brownian motion with \rt{quorum sensing based} interactions}
\label{sec:abp}

\begin{figure}[t]
\centering
\includegraphics[trim={0 4.3cm 0 2cm},clip,scale=0.3]{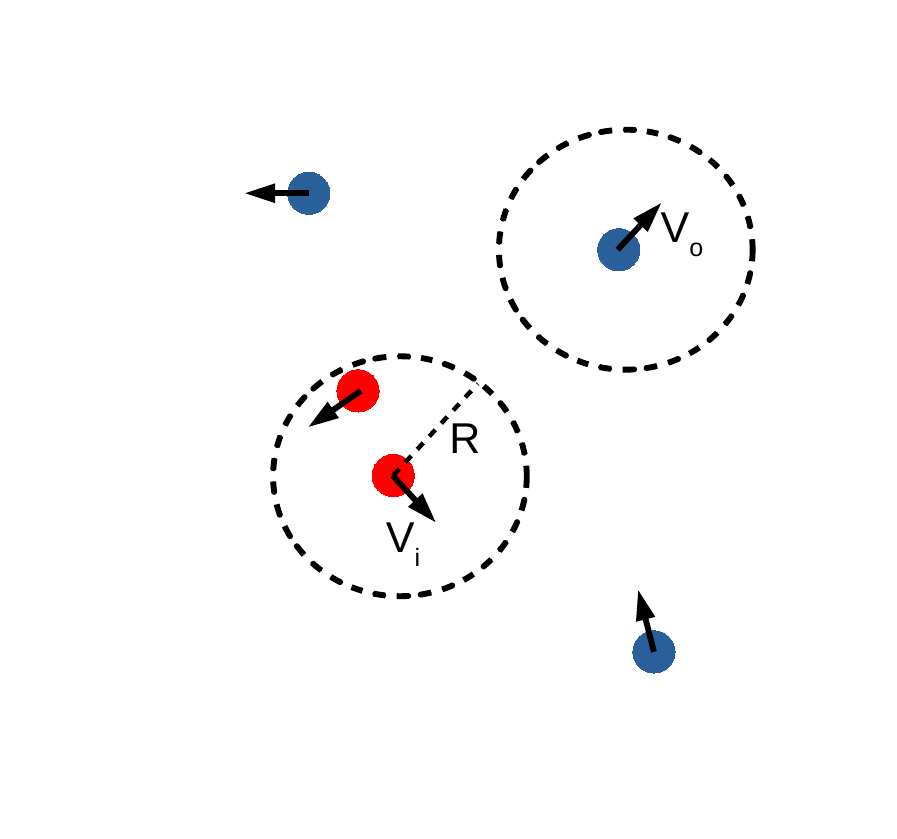}
\caption{\rt{Quorum sensing based} interaction. Each particle senses whether there are other particles within an interaction radius, $R$. If there is at least one other particle within this range, the particle (colored red) moves at velocity $V_i$. Otherwise, it (colored blue) moves at velocity $V_o$. The particles do not interact physically, but only via sensing-based rules, and consequently may be considered point particles. Therefore the size of the particles in this and subsequent figures has no physical meaning.} 
\label{fig-carton}
\end{figure}

We consider particles performing active Brownian motion, such that the dynamics for each particle in three dimensions (3D) is governed by the following set of equations of motion~\cite{bechinger2016},
\begin{subequations}
\label{eq-abp1}
\begin{eqnarray}
    \frac{dx_j}{dt}&=&v_j(t) \cos{\theta_j(t)} \cos{\phi_j(t)}, \\
    \frac{dy_j}{dt}&=&v_j(t) \cos{\theta_j(t)}\sin{\phi_j(t)}, \\
    \frac{dz_j}{dt}&=&v_j(t) \sin{\theta_j(t)}, \\
    \frac{d\theta_j}{dt}&=&\sqrt{2\mathcal{D}}\eta_j(t). \\
    \frac{d\phi_j}{dt}&=&\sqrt{2\mathcal{D}}\xi_j(t).
\end{eqnarray}
\end{subequations}
Here, $x_j$, $y_j$ and $z_j$ are the position coordinates of particle $j$, and $v_j$ is the magnitude of its velocity, i.e. the self propulsion speed of particle $j$.  $\phi_j$ is the angle subtended by the velocity vector with the $x-y$ plane while  $\theta_j$ is the angle subtended with the $x$ axis by the projection of the velocity vector on the $x-y$ plane. They describe the direction of motion of the particle and perform independent rotational diffusion with $\mathcal{D}$, the rotational diffusion constant, and $\xi_j(t)$ and $\eta_j(t)$  are uncorrelated Gaussian white noise with zero mean and unit variance, i.e., $\langle \xi_j(t) \xi_k(t') \rangle = \delta_{jk} \delta(t-t')$, $\langle \eta_j(t) \eta_k(t') \rangle = \delta_{jk} \delta(t-t')$ and $\langle \xi_j(t) \eta_k(t') \rangle =0$. We also consider a two-dimensional (2D) variant of our model, in which we set $z_j \equiv 0$ and $\theta_j \equiv 0$, such that the motion is only in the $x-y$ plane.

While $\theta_j$ and $\phi_j$ change instantaneously, $1/\mathcal{D}$ sets a persistence time and $v_j/\mathcal{D}$ a persistent length for directional motion, thereby allowing the artificial particles governed by Eq.~(\ref{eq-abp1}) to mimic the dynamics of natural organisms for which one expects a non-instantaneous change of orientation. We consider only orientational fluctuations in Eqs.~(\ref{eq-abp1}) and exclude translational fluctuations because our focus in this study is to propose a minimal model of ABP that can give rise to  emergent clustering properties. See~\cite{weber2016binary, mccarthy2023demixing} for the role of translational diffusivity in clustering of binary mixtures. In our case we expect orientational fluctuations to have a more significant effect, since translational fluctuations may be coupled to the propulsion (translation) of the particles. This is particularly relevant considering the potential application of this model to artificial robotic swimmers, where translational fluctuations are not expected, but orientational fluctuations may be imposed by mechanical design to realize specific outcomes.

We consider the \rt{quorum sensing based}, virtual interaction between particles as depicted in Fig.~\ref{fig-carton}: if a particle senses at least one other particle within an interaction radius, $R$, the magnitude of its velocity is $V_i$ (red), whereas it is $V_o$ if there are no particles within this range (blue). Note that we use the subscript $i$  to denote velocities of particles when there is at least one neighbor present within the interaction radius, whereas we use the subscript $o$ to denote velocities of particles when there is no neighbor present within the interaction radius. These should not be confused with the particle indices. We distinguish between two types of interactions: (a) \textit{social} behavior when $V_i<V_o$, i.e. a particle slows down when it senses other particles within $R$, and (b) \textit{anti-social} behavior when $V_i>V_o$, i.e. a particle moves faster when it senses other particles within~$R$. The particles do not interact physically, but only via these sensing-based rules, and consequently may be considered point particles. Therefore the size of the particles in Fig.~\ref{fig-carton} and subsequent figures has no physical meaning. See Ref.~\cite{zhou2023quorum} for an investigation of the effect of particle size on the clustering of ABP with perception based interactions. 

\begin{figure*}[t]
\includegraphics[scale=0.44]{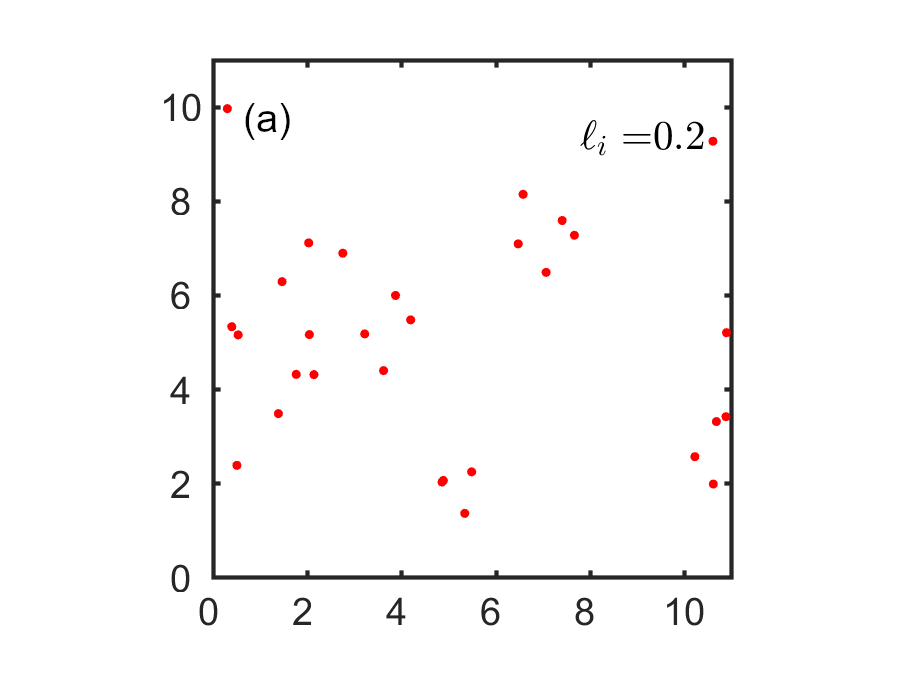}
\hspace{-1.5cm}
\includegraphics[scale=0.44]{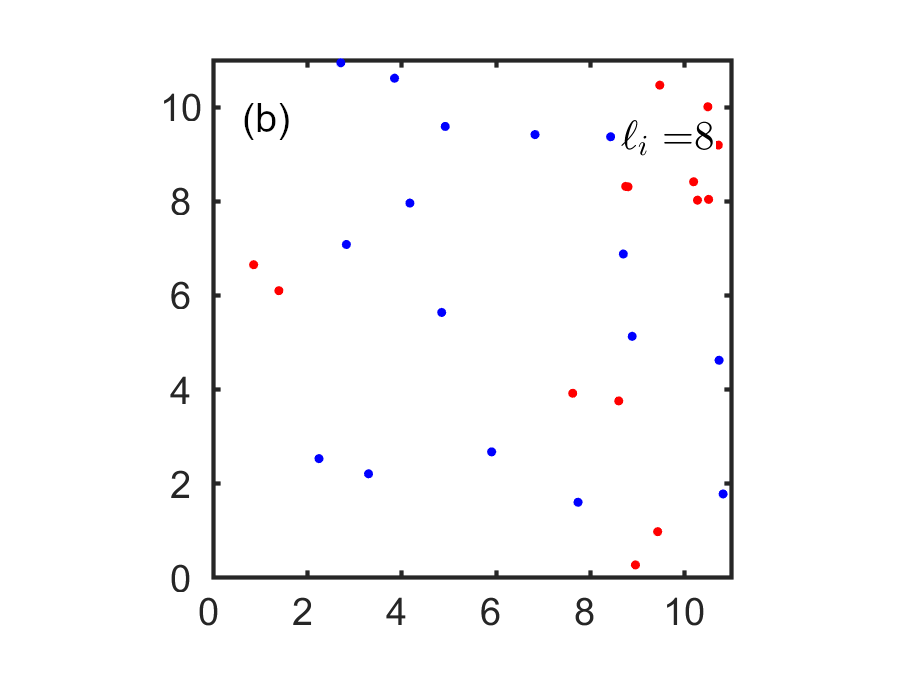}
\hspace{-1.5cm}
\includegraphics[scale=0.44]{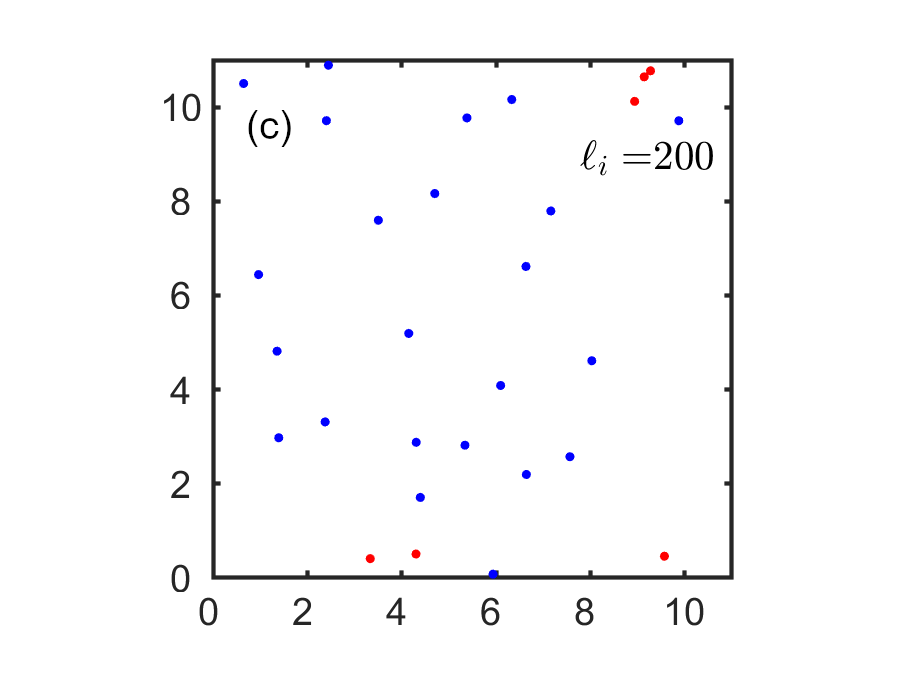}
\caption{Representative realization of the steady-state positions of the particles in a 2D simulation with $N=30$ particles. Positions are normalized by the interaction radius $R$, and the model parameters are $\ell_o=20$, $\ell_N=2$, with (a) $\ell_i=0.2$, (b) $\ell_i=8$ and (c) $\ell_i=200$. (a) and (b) correspond to social behavior i.e. $\ell_i<\ell_o$, while (c) corresponds to anti-social behavior, i.e. $\ell_i>\ell_o$.}
\label{fig-spos}
\end{figure*}

\section{Numerical Simulations}
\label{sec:results}

We performed numerical simulations with $N$ particles in a square box of size $L \times L$ for 2D motion or a cubic box of size $L \times L \times L$ for 3D motion, both under periodic boundary conditions. The numerical integration of the set of Eqs.~(\ref{eq-abp1}) is performed by introducing a short simulation time-step $\delta t$ such that  starting from random initial conditions,  the coordinates after $m+1$ steps are obtained from those after step $m$ using
\begin{subequations}
\label{eq-abpsim}
\begin{eqnarray}
    \Delta \theta_{j,m}&=&W_{j,m}\sqrt{2\mathcal{D}\delta t},\\
    \Delta \phi_{j,m}&=&Q_{j,m}\sqrt{2\mathcal{D}\delta t},\\
    x_{j,m+1}&=&x_{j,m}+v_{j,m} \cos{\left(\theta_{j,m}+\frac{\Delta \theta_{j,m}}{2}\right)} \nonumber \\
    &&\cdot\cos{\left(\phi_{j,m}+\frac{\Delta \phi_{j,m}}{2}\right)} \delta t ,\\
    y_{j,m+1}&=&y_{j,m}+v_{j,m} \cos{\left(\theta_{j,m}+\frac{\Delta \theta_{j,m}}{2}\right)}  \nonumber \\
    &&\cdot\sin{\left(\phi_{j,m}+\frac{\Delta \phi_{j,m}}{2}\right)} \delta t ,\\
    z_{j,m+1}&=&z_{j,m}+v_{j,m} \sin{\left(\theta_{j,m}+\frac{\Delta \theta_{j,m}}{2}\right)} \delta t,\\
    \theta_{j,m+1}&=&\theta_{j,m}+\Delta \theta_{j,m}, \\
     \phi_{j,m+1}&=&\phi_{j,m}+\Delta \phi_{j,m},
\end{eqnarray}
\end{subequations}
where $W_{j,m}$ and $Q_{j,m}$ are Gaussian distributed random numbers with zero mean and unit variance. We verified that a numerical time step of $\delta t=10^{-3}$ is small enough for the results to be insensitive to its value. Each realization of the simulation is run for a total time of at least $T=10^7$ steps. We verified that all the quantities that we measure have converged to their steady values after this time. See Figs.~\ref{fig-gr-steady} and~\ref{fig-clust-steady} for the relaxation of the system to its steady state.

The equations of motion have the following dimensional parameters: $R$, $V_i$, $V_o$ and $\mathcal{D}$. We measure distances in units of $R$ and measure time in units of $1/\mathcal{D}$. The two speeds $V_i$ and $V_o$ introduce two dimensionless length scales: $\ell_i=V_i/(R \mathcal{D})$ and $\ell_o=V_o/(R \mathcal{D})$. These correspond to the dimensionless persistence lengths inside and outside the interaction range, respectively, both measured in units of the interaction range. Finally, the number density of particles in the system introduces another length scale $L/N^{1/2}$ in 2D and $L/N^{1/3}$ in 3D, which give the third dimensionless length, $\ell_N=L/(N^{1/2}R)$ and $\ell_N=L/(N^{1/3}R)$ in two and three dimensions, respectively.

Figure~\ref{fig-spos} shows results from 2D simulations and  depicts particular realizations of the positions of the particles in the steady state corresponding to social (Fig.~\ref{fig-spos}a and Fig.~\ref{fig-spos}b) and anti-social behavior (Fig.~\ref{fig-spos}c) for fixed $\ell_o=20$ and $\ell_N=2$. A particle is colored red whenever it has at least one more particle within the interaction radius, otherwise it is colored blue. Figure~\ref{fig-spos}a shows that in case of $\ell_i \ll \ell_o$, the particles cluster into rather big and well-spread groups, with almost every particle having at least another particle within the interaction radius. Conversely, for larger values of  $\ell_i$ the particles do not form big clusters as evident from the mixture of red and blue particles, as shown in Fig.~\ref{fig-spos}b. The anti-social case with $\ell_i \gg \ell_o$ -- as seen in Fig.~\ref{fig-spos}c -- results in a steady state where almost every  particle has a nearest neighbor at a distance slightly larger than the interacting radius. The simulations shown here are with $N=30$ particles. Figure~\ref{fig-spos500} in Appendix~\ref{sec:appendix} shows similar behavior for a system of $N=500$ particles. Based on the quantitative system-size analysis detailed below we concluded that a system size of $N=30$ particles is large enough for studying the physical phenomena that we focus on.

\begin{figure*}[t]
\vspace{-0.4cm}
\includegraphics[width=0.8\columnwidth]{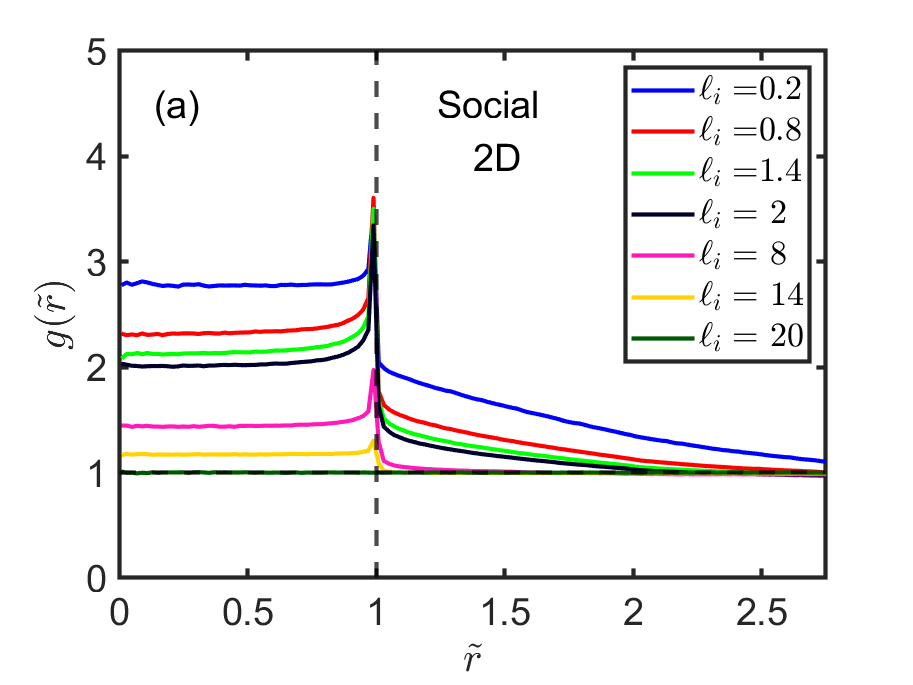}
\includegraphics[width=0.8\columnwidth]{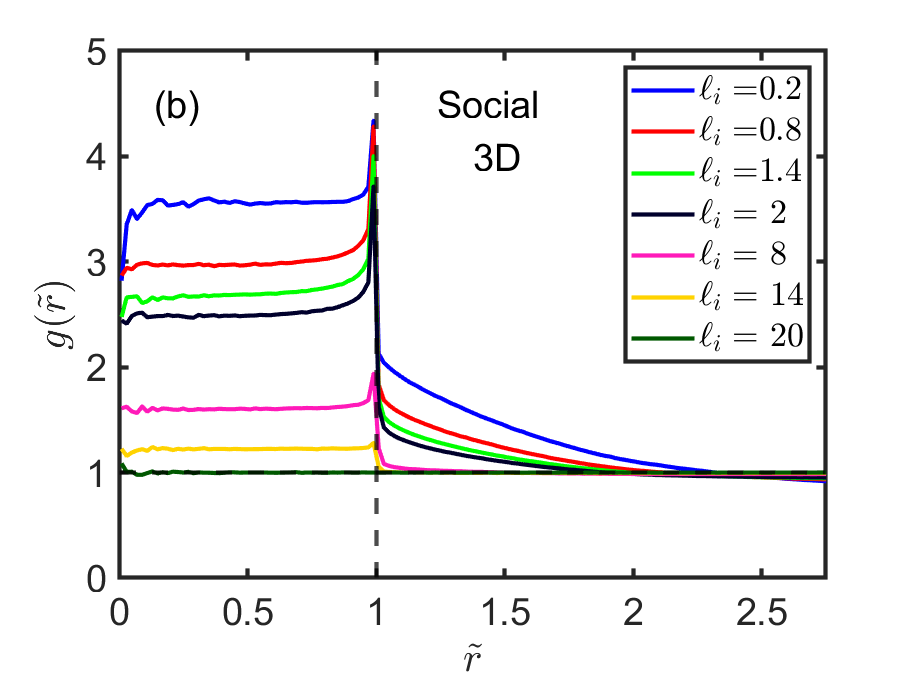}\\
\includegraphics[width=0.8\columnwidth]{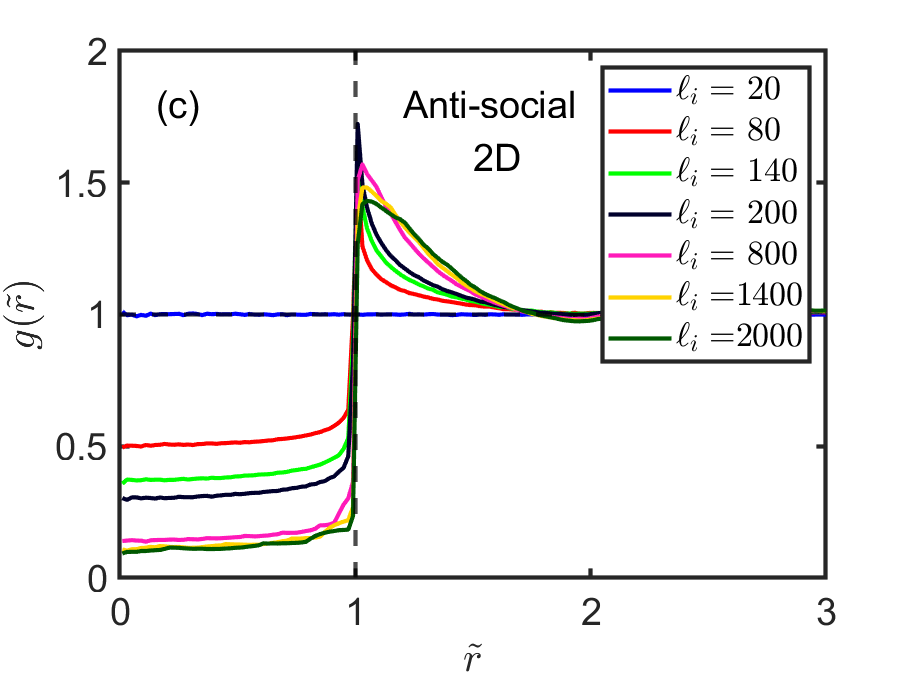}
\includegraphics[width=0.8\columnwidth]{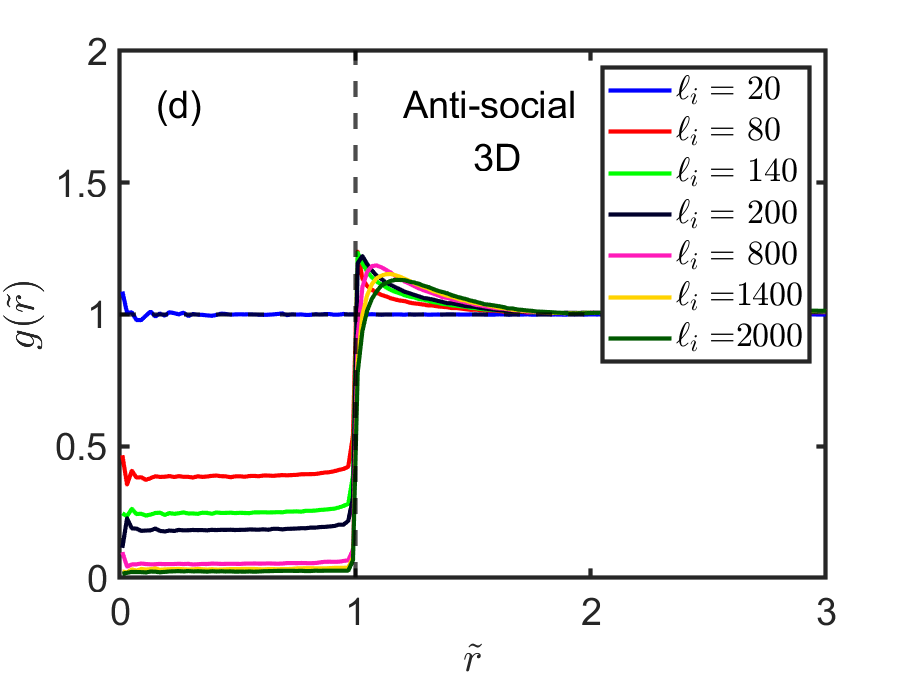}
\vspace{-0.2cm}
\caption{The radial distribution function $g(\tilde{r})$ vs. the normalized distance between particles $\tilde{r}=r/R$ in the case of 2D (a $\&$ c) and 3D (b $\&$ d) motion. (a) $\&$ (b) social interaction, i.e. $\ell_i<\ell_o$, while (c) $\&$ (d) anti-social interaction, i.e. $\ell_i>\ell_o$. The dashed line corresponding to $\tilde{r}=1$ shows the interaction radius. Here, $N=30$, $\ell_o=20$, $\ell_N=2$, and $\ell_i$ is indicated in the legend. Note that in 3D $\ell_N=L/(RN^{1/3})$ whereas in 2D $\ell_N=L/(RN^{1/2})$.}
\label{fig-gr}
\end{figure*}

To quantify these observations, we measured  the radial distribution function $g(r)$, which we describe next~\cite{chaikin1995,burgot2017notion}. The one-point density function, $\rho^{(1)}(\mathbf{r}')$ and two-point density function, $\rho^{(2)}(\mathbf{r}',\mathbf{r}'')$ are required for this. $\rho^{(1)}(\mathbf{r}')d\mathbf{r}'$ gives the probability that a particle will be found in volume $d\mathbf{r}'$ around position $\mathbf{r}'$. The integration of the one-point density function over the entire volume, $V$,  gives the total number of particles $N$. In our case, the medium is statistically homogeneous and isotropic and therefore,
\begin{equation}
\label{eq-numdensi}
    \rho^{(1)}(\mathbf{r}')=\frac{N}{V}\equiv \rho_N,
\end{equation}
where $\rho_N$ is the average number density.

The term $\rho^{(2)}(\mathbf{r}',\mathbf{r}'')d\mathbf{r}'d\mathbf{r}''$ is the joint probability to find one particle in $d\mathbf{r}'$ at $\mathbf{r}'$ and one particle in $d\mathbf{r}''$ at $\mathbf{r}''$. The pair correlation function $g(\mathbf{r}',\mathbf{r}'')$ is defined from
\begin{equation}
\label{eq-pair-corr}
    \rho^{(2)}(\mathbf{r}',\mathbf{r}'')=\rho^{(1)}(\mathbf{r}')\rho^{(1)}(\mathbf{r}'')g(\mathbf{r}',\mathbf{r}'').
\end{equation}
Again, due to homogeneity and isotropy, the two-point density depends only on the scalar distance $r=|\mathbf{r}'-\mathbf{r}''|$, such that we can write Eq.~(\ref{eq-pair-corr}) as
\begin{equation}
    \label{eq-gr}
    \rho^{(2)}(r)=\rho_N^2g(r),
\end{equation}
where we used Eq.~(\ref{eq-numdensi}). $g(r)$ appearing in Eq.~(\ref{eq-gr}), i.e. the pair correlation function in case of homogeneous and isotropic fluid, is called the radial distribution function~\cite{chaikin1995,burgot2017notion}. In an uncorrelated system, such as an ideal gas, the probability of finding a particle at any position is uniform and independent of the positions of the other particles. In this case $g(r)$ does not depend on $r$, and $g(r) \equiv 1$. Interparticle interactions can lead to spatial correlations that become expressed in $g(r)$. While $g(r)>1$ represents positive correlations and is a signature of effective attraction, in the case of repulsion negative correlations develop, and $g(r)<1$.

The integral of $\rho_Ng(r)$ over the entire volume, $V$ gives the total number of particles, $N$. This results in the normalization condition,
\begin{equation}
    \label{eq-normalizeg}
    \frac{1}{V}\int_V g(r)dV=1,
\end{equation}
where $dV=2\pi r dr$ in 2D whereas  $dV=4 \pi r^2 dr$ in 3D. The integral of $\rho_Ng(r)$ over a volume element $dV$ gives the number of particles in that volume element~\cite{chaikin1995}. Thus $g(r)$ can be obtained from the simulations by counting the number of particles in a small volume $dV$ at a separation $r$ from a particle at the origin and dividing by $\rho_NdV$.
 
We measured $g(r)$ in the steady state and present our results in terms of the dimensionless inter-particle distance $\tilde{r}=r/R$. Figure~\ref{fig-gr} shows how $g(\tilde{r})$ varies with $\tilde{r}$, in both 2D and 3D simulations. For social behavior, i.e. $\ell_i<\ell_o$, we see that $g(\tilde{r})>1$ for $\tilde{r}<1$, exhibiting attractive interaction between the particles up to the interaction radius. For $\tilde{r}>1$, $g(\tilde{r})$ decays to one, characterizing the lack of correlations at large separation between the particles. We also observe a distinction between two types of decay of $g(\tilde{r})$ for $\tilde{r}>1$; Namely, when $\ell_i \ll \ell_o$, $g(\tilde{r})$ exhibits a slow decay with $\tilde{r}$, while it exhibits a sharp decay to $g(r)=1$ otherwise. These distinct behaviors can be understood by looking at representative positions of the particles in the steady-state, as shown in 2D in Fig.~\ref{fig-spos}a,b. The steady state positions corresponding to the other 2D cases in Fig.~\ref{fig-gr}a are presented in Fig.~\ref{fig-social-pos} in Appendix~\ref{sec:appendix}, and videos showing the time-evolution of the particles in the steady state corresponding to all the cases shown in Fig.~\ref{fig-gr}a are presented as supplementary files. The  clustering of particles in the steady state for $\ell_i \ll \ell_o$ -- as seen for instance in Fig.~\ref{fig-spos}a -- results in significant contributions to $g(\tilde{r})$ from a range of inter-particle distances $\tilde{r}>1$, thereby leading to a slow decay of $g(\tilde{r})$ for $\tilde{r}>1$. Conversely, due to the lack of clustering for larger values of  $\ell_i$ -- as shown for instance in Fig.~\ref{fig-spos}b -- there are not enough contributions to  $g(\tilde{r})$ from inter-particle distances of $\tilde{r}>1$ resulting in  $g(\tilde{r})$ transitioning rather sharply to $g(r)=1$. 
 
In case of anti-social behavior, i.e. $\ell_i>\ell_o$, we see that $g(\tilde{r})<1$ for $\tilde{r}<1$, exhibiting negative correlations, or an effective repulsive interaction between the particles up to the interaction radius. Remarkably, for $\tilde{r}>1$, there is a regime in which $g(\tilde{r})>1$ corresponding to attractive interaction even for the anti-social case. Finally, $g(\tilde{r})$ decays to one, characterizing the lack of interactions at large separation between the particles. This demonstration of attractive interaction, as displayed by $g(\tilde{r})>1$, results from the accumulation of particles at distances slightly greater than $\tilde{r}=1$ from their nearest neighbors.  Upon detection of a neighbor within the interaction radius, a particle takes velocity $v=V_i>V_o$ and therefore soon finds itself outside the interaction range of that neighbor. When there are no neighbors within the interaction radius, the particle chooses $v=V_o<V_i$. For an ensemble of particles, this results in a steady state where almost every  particle has a nearest neighbor at a distance slightly larger than the interacting radius. This is seen, for example, for the blue particles in Fig.~\ref{fig-spos}c. This is also observed for all the cases with $\ell_i \gg \ell_o$ shown in Fig.~\ref{fig-asocial-pos} in Appendix~\ref{sec:appendix} where we show steady state positions corresponding to all the  cases of Fig.~\ref{fig-gr}c except the one considered in Fig.~\ref{fig-spos}c.  Consequently, there is a significant contribution to $g(\tilde{r})$ from inter-particle distances slightly larger than $\tilde{r}=1$, manifested as $g(\tilde{r})>1$  observed in Fig.~\ref{fig-gr}c. Videos showing the time-evolution of the particles in the steady state corresponding to all the cases shown in Fig.~\ref{fig-gr}c are presented as supplementary files. 

Interestingly, the results in 3D (Figs.~\ref{fig-gr}b and \ref{fig-gr}d) exhibit the same features as in 2D (Figs.~\ref{fig-gr}a and \ref{fig-gr}c), with some quantitative differences. Specifically, due to geometric reasons, the fraction of volume inside the interaction sphere out of the total volume of 3D space is smaller than the fraction of area inside the interaction disc out of the total area of 2D space. Therefore, the deviations within the interaction range of the pair correlation function from the uncorrelated value of 1 are larger in 3D than in 2D, both for social and for anti-social behavior. Considering the similar behavior in two and in three dimensions, we will restrict ourselves to the investigations of 2D motion in the rest of the article. 

\begin{figure}[]
\centering
\includegraphics[width=0.9\columnwidth]{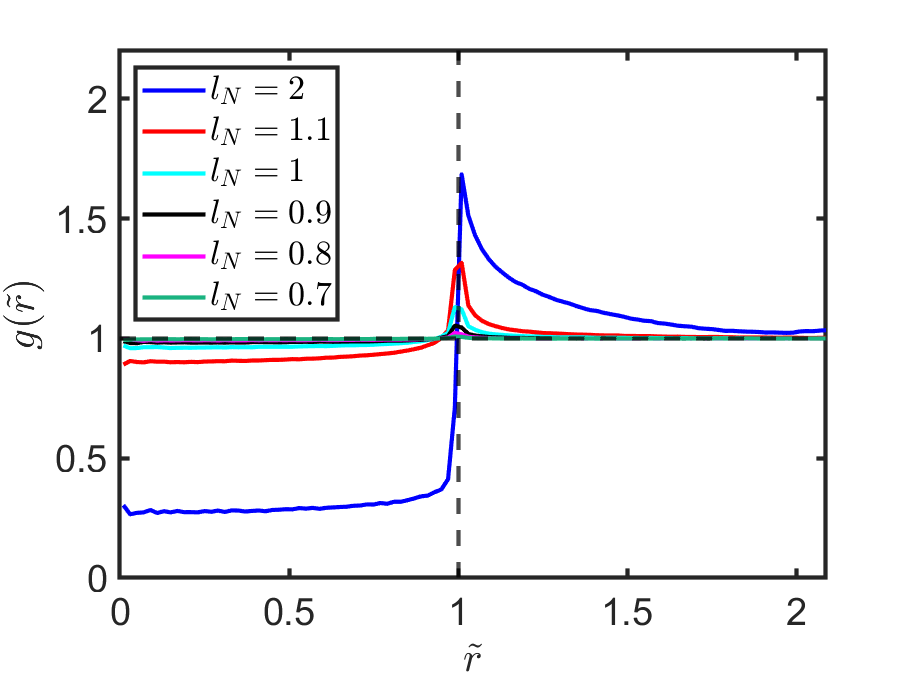}
\caption{The radial distribution function $g(\tilde{r})$ vs. the normalized distance between particles $\tilde{r}=r/R$ for 2D anti-social interaction with $\ell_i=200$, $\ell_o=20$ and different $\ell_N$ as shown in the legend. $L$ and $R$ are fixed and $N$ takes the values $(10,30,40,50,60,70)$, such that different values of $\ell_N$ correspond to different number densities.  The dashed line, corresponding to $\tilde{r}=1$, shows the interaction radius. }
\label{fig-density}
\end{figure}

To study the density dependence, Fig.~\ref{fig-density} presents how $g(\tilde{r})$ varies with $\tilde{r}$ in the 2D anti-social case with $\ell_i=200$, $\ell_o=20$ and different $\ell_N$ as shown in the legend. With $L$ and $R$ fixed, $N=[10, 30, 40, 50, 60, 70]$ is varied (top to bottom in the legend) such that different values of $\ell_N$ correspond to different number densities, $\rho_N=N/L^2$. Thus, larger values of $\ell_N$ corresponds to lower densities. Figure~\ref{fig-density} shows that the attractive interaction in case of anti-social behavior is substantial only when the number density of particles is low.

\begin{figure}
\centering
\includegraphics[width=0.9\columnwidth]{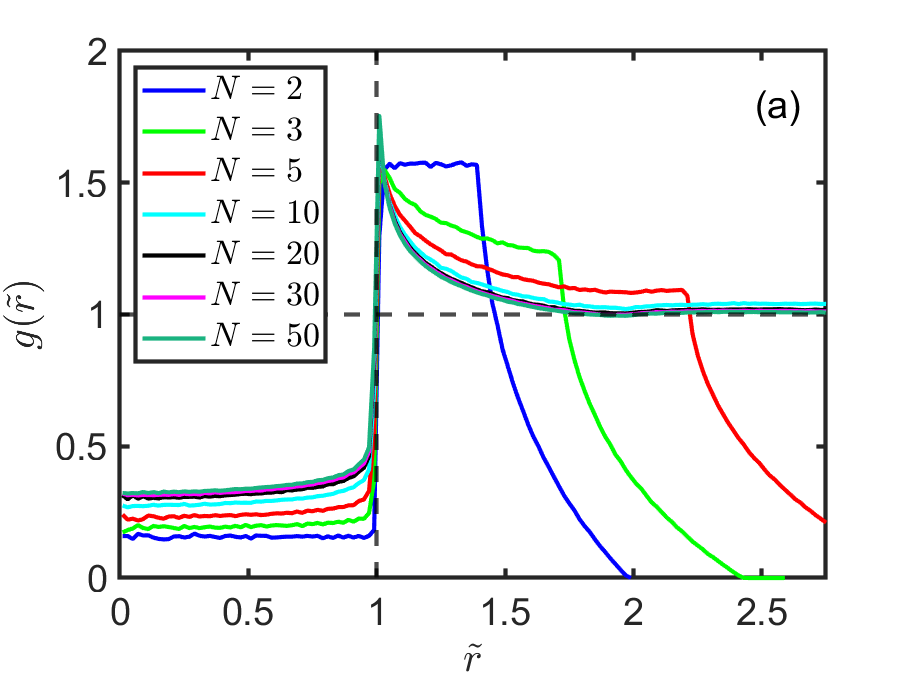}\\
\includegraphics[width=0.9\columnwidth]{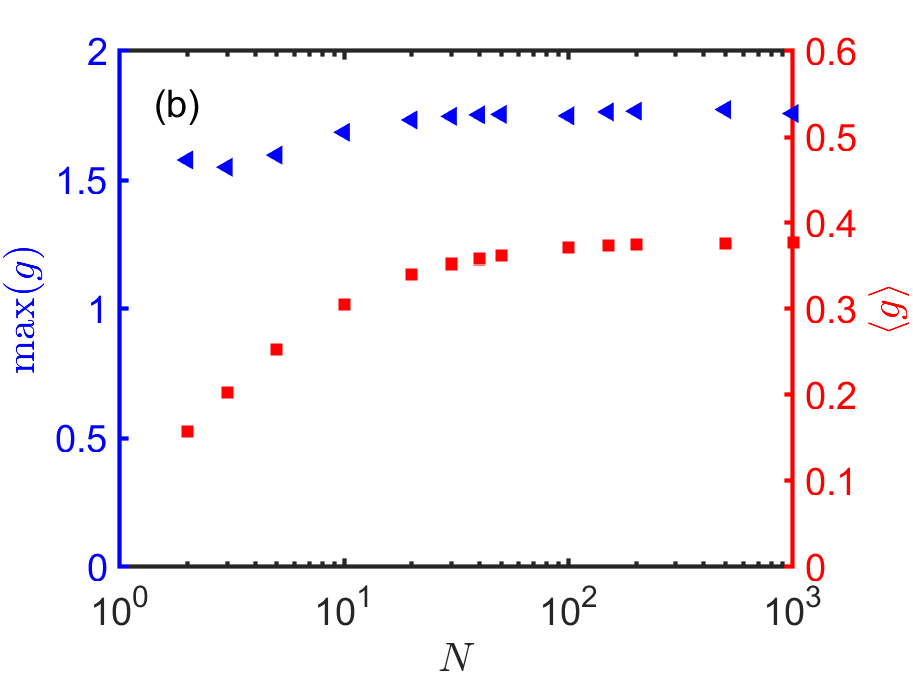}
\caption{Effect of the number of particles on the radial distribution function $g(\tilde{r})$ for the anti-social case in 2D. Here, $\ell_i=200$, $\ell_o=20$, and  $\ell_N=2$. (a) $g(\tilde{r})$ vs. $\tilde{r}$ for different~$N$. (b) Average $\langle g \rangle$ vs. $N$ (red squares, right axis), where the average is over the range $\tilde{r}<1$; $\max(g)$ vs. $N$ (blue triangles, left axis).} 
\label{fig-deffect}
\end{figure}

\begin{figure}[t]
\includegraphics[width=0.9\columnwidth]{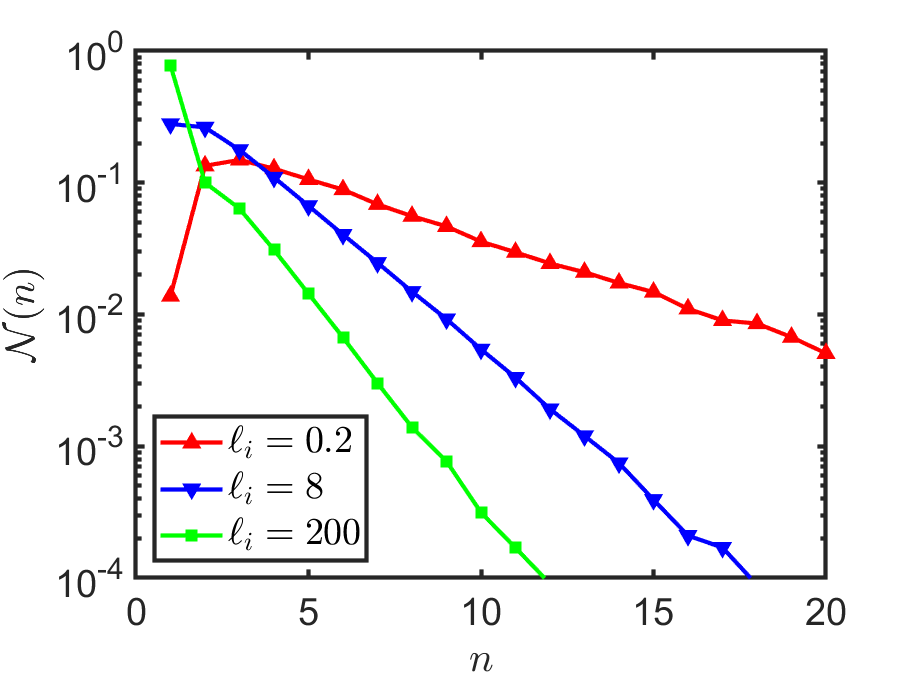}
\caption{The cluster size distribution $\mathcal{N}(n)$ in 2D vs. the cluster size $n$ with $N=500$, $\ell_N=2$, $\ell_o=20$ and different values of $\ell_i$, as indicated in the legend.} 
\label{fig-clusdist}
\end{figure}

\begin{figure}[]
\includegraphics[width=0.9\columnwidth]{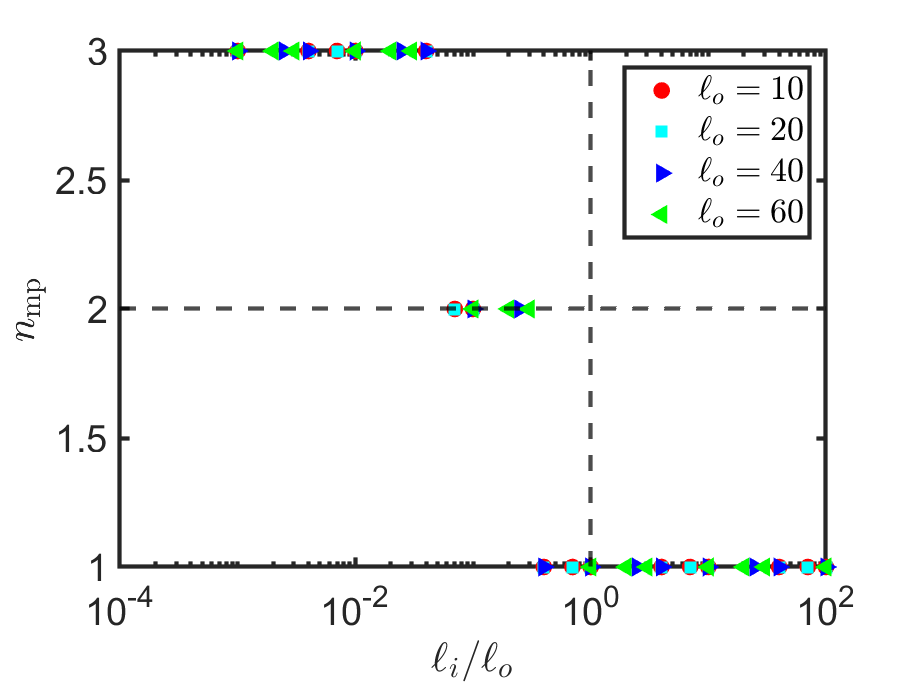}
\caption{Most probable cluster size $n_\mathrm{mp}$ vs. $\ell_i/\ell_o$ with fixed $\ell_N=2$ and $N=500$ in 2D.}
\label{fig-mpn}
\end{figure}

Given the non-reciprocal, many-body interactions of our model, one may ask how many particles are needed to observe the emergent attractive interactions in case of anti-social behavior. Figure~\ref{fig-deffect}a shows that even $N=2$ particles are enough to exhibit such attractive interactions. We observe that for $2\leq N<10$ particles, there is a regime such that $g(\tilde{r})>1$ denoting attractive interaction even with anti-social rules. As expected, $g(\tilde{r})<1$ for $\tilde{r}<1$, exhibiting the repulsive interaction between the particles up to the interaction radius. The saturation of $g(\tilde{r})$ to one, signifying the non-interacting regime is not visible because of the finite size of the simulation box, which becomes important for $\tilde{r}>L/(R\sqrt{2})$ and results in the decay of $g(\tilde{r})$ to zero. However, with $N \geq 10$ particles we do observe all three effects, namely, $g(\tilde{r})<1$ for $\tilde{r}<1$, $g(\tilde{r})>1$ for $\tilde{r}>1$, and finally $g(\tilde{r})$ decays to one for $\tilde{r} \gg 1$. We note that the results presented in Fig.~\ref{fig-deffect} are at fixed low number density corresponding to $\ell_N=2$.

To quantitatively test the convergence of the simulation results as system size is increased, we consider $\langle g \rangle$, the average correlations within the interaction radius $\tilde{r}<1$ defined as
\begin{equation}
    \langle g \rangle= \int_{0}^{1}g(\tilde{r})d\tilde{r}.
\end{equation} 
This quantifies the effective attraction or repulsion between particles. We also consider the maximal correlation $\max(g)$, which here appears slightly beyond the interaction range. Figure~\ref{fig-deffect}b demonstrates that already $N=10$ behaves qualitatively similarly to larger systems, and that quantitative convergence starts around $N=100$. This justifies the use of $N=30$ for most of the study presented above, and the conclusion that even $N=10$ could be enough to see these qualitative results.

To investigate the clustering, as an alternative to $g(r)$,  we present in Fig.~\ref{fig-clusdist} the cluster-size distribution $\mathcal{N}(n)$ which gives the average fraction of particles in clusters of size $n$ particles, and is defined as  
~\cite{negi2022emergent,peruani2006nonequilibrium,theers2018clustering}
\begin{equation}
\mathcal{N}(n)=\frac{1}{N_N}nP(n),
\label{eq-Nclustdist}
\end{equation}
where $P(n)$ is the number of clusters of size $n$, and the normalization constant $N_N$ is set such that $\sum_n\mathcal{N}(n)=1$. A particle belongs to a specific cluster if it is separated by a distance less than the interaction radius $R$ from at least one member of the cluster.
%We chose $N=500$ for all the analyses of cluster size statistics (Figs.~\ref{fig-clusdist}, \ref{fig-mpn} and \ref{fig-ell}b), in contrast to $N=30$ for all the analyses of $g(r)$ unless stated otherwise. This is because the clusters are distinguishable even with a visible inspection of the steady state positions with $N=500$ particles as seen in Fig.~\ref{fig-spos500} in Appendix~\ref{sec:appendix}. This figure also shows that the steady state positions of the particles demonstrate similar behavior as seen in Fig.~\ref{fig-spos}. Considering the similar behavior of the cluster size distribution for N=30 and N=500, we checked for the convergence of the cluster size statistics to the steady state in Fig.~\ref{fig-clust-steady} with $N=30$.} 

Figure~\ref{fig-clusdist} shows the cluster size distribution $\mathcal{N}$ vs. the cluster size $n$, with the parameters corresponding to Fig.~\ref{fig-spos}, but with $N=500$ particles and $L$ chosen such that $\ell_N=2$. Remarkably, Fig.~\ref{fig-clusdist} shows an exponential decay of $\mathcal{N}$ vs. $n$ for $n>n_{\mathrm{mp}}$ where $n_{\mathrm{mp}}$ denotes the most probable cluster size.  Indeed,  in case of $\ell_i \gg \ell_o$, a cluster size $n=1$ has a probability of very close to unity, highlighting that most of the particles do not have any neighbor within the interaction radius.

Figure~\ref{fig-mpn} shows the most probable cluster size $n_\mathrm{mp}$ vs. $\ell_i/\ell_o$ with fixed $\ell_N=2$ and each curve corresponding to fixed $\ell_o$ as shown in the legend.  Interestingly, the results depend not on the magnitudes of $\ell_i$ and $\ell_o$, but rather their ratio.  The most probable cluster size for the case $\ell_i \ll \ell_o$ is $n_\mathrm{mp}$ = 3, whereas when $\ell_i$ is comparable or greater than $\ell_o$ the most probable cluster size is $n_\mathrm{mp}=1$. Thus, even in the very social case, the clusters do not merge to form a single giant cluster. In the anti-social case, as expected, each particle prefers to remain isolated thereby resulting in $n_\mathrm{mp}=1$.

\begin{figure}[b]
\centering
\includegraphics[width=0.9\columnwidth]{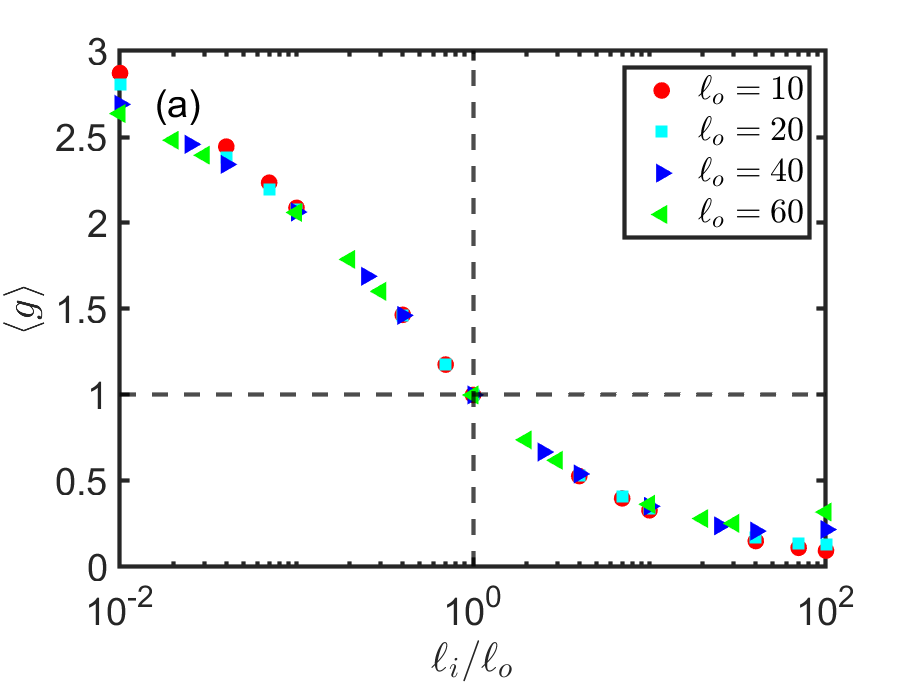}\\
\includegraphics[width=0.9\columnwidth]{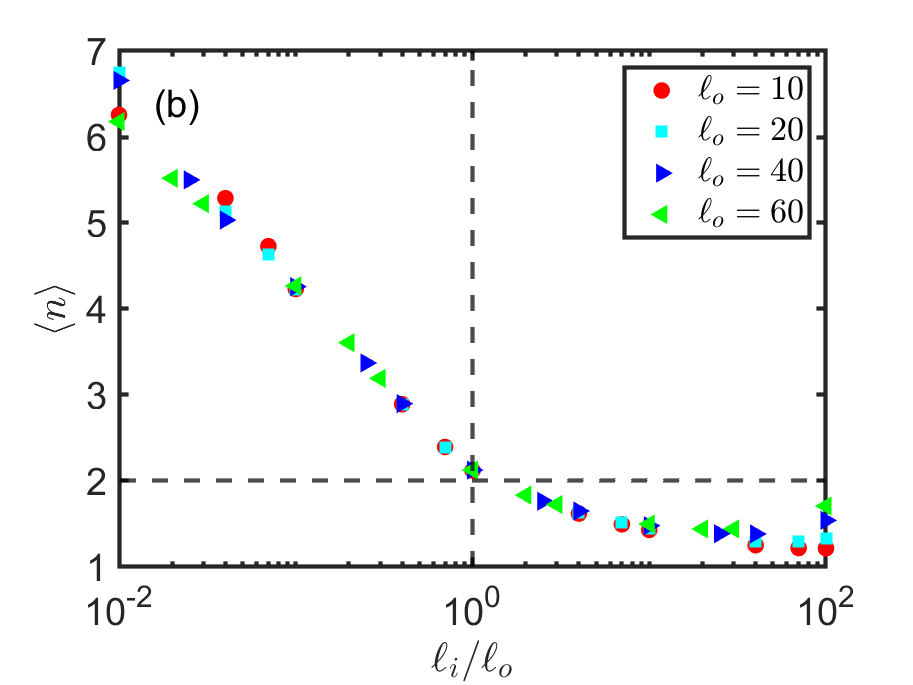}
\caption{(a) Average correlation $\langle g \rangle$ and (b) average cluster size $\langle n \rangle$ vs. $\ell_i/\ell_o$, with fixed $\ell_N=2$ in 2D. The vertical dashed line shows the transition from social to anti-social behavior. The correlation is averaged over the interaction range $\tilde{r}<1$, and the horizontal dashed line shows the value of $\langle g \rangle = 1$ for an uncorrelated system. (a) is with $N=30$ particles and (b) with $N=500$.}
\label{fig-ell}
\end{figure}

Figure~\ref{fig-ell}a shows how the average $\langle g \rangle$ -- averaged over the interaction range $\tilde{r}<1$ -- varies as a function of $\ell_i/\ell_o$. Here, $\ell_N=2$ and different symbols correspond to fixed values of $\ell_o$ as shown in the legend. Remarkably, the results collapse to a master curve implying that the observed trends depend not on the magnitudes of $\ell_i$ and $\ell_o$, but rather their ratio. The observed $\langle g \rangle$ clearly shows the transition from $\langle g \rangle >1$ for $\ell_i/\ell_o <1$ to $\langle g \rangle <1$ for $\ell_i/\ell_o >1$. This trend corresponds to the transition from positive correlations in case of social behavior to negative correlations in case of anti-social behavior. Figure~\ref{fig-ell}b shows how the average cluster size $\langle n \rangle$ -- averaged over the distribution $\mathcal{N}(n)$, i.e. $\langle n \rangle=\sum_{n}n\mathcal{N}(n)$ -- varies as a function of $\ell_i/\ell_o$. It exhibits collapse to a master curve similar to that seen in Fig.~\ref{fig-ell}a for $\langle g \rangle$. Here we observe a transition from $\langle n \rangle >2$ for $\ell_i/\ell_o <1$ to $\langle n \rangle<2$ for $\ell_i/\ell_o >1$. This trend mimics the trend seen for $\langle g \rangle$, namely the different behavior resulting from positive correlations in case of social behavior and negative correlations in case of anti-social behavior.

\section{Discussion}
\label{sec:discuss}

Our simple model with short-ranged \rt{quorum sensing based} interactions generates collective motion of multiple swimmers. Tuning interaction parameters such as the velocities of the swimmers results in two main behaviors: clustering or anti-clustering. Yet, additional phenomena emerge from the model. For example, even in cases in which a group of swimmers tends to disperse, swimmers may cluster. Such behavior is observed even for just two active swimmers. The simplicity of the model, together with the resemblance of the virtual interactions to the way group members communicate in nature, makes this model a powerful tool for engineering bio-inspired coordinated groups of robots. Positive correlations emerge in the case of anti-social behavior not only in two dimensions but also in three dimensions although it is less pronounced in the latter case due to the larger physical space that the particles are free to move in.

The study conducted here should be contrasted with MIPS which has been the subject of several recent investigations~\cite{cates2008,cates2015motility, marchetti2012, merrigan2020, iyer2023dynamics,farrell2012pattern,paoluzzi2020information}.  In our model we have many-body and non-reciprocal interactions, and the clustering that we observe is not a result of the motility  but rather as a consequence of how motility is affected by the interactions, namely, how a particular velocity is chosen by a particle depending on whether it detects a neighbor within the interaction radius or not.  Moreover, MIPS often leads to phase separation, in which eventually the entire system separates into one giant cluster in a condensed phase and the rest of the system is in a dilute gas phase~\cite{cates2015motility}. Yet, arrested coarsening where multiple clusters are formed have also been reported~\cite{van2019interrupted,gonnella2015motility}. In our case, we have multiple clusters that do not coarsen to reach the entire system size (see Fig.~\ref{fig-clusdist}). Another significant difference of our system from MIPS is that the latter is an effect of social interaction and is stronger for dense systems, whereas we observe  clustering also  in case of anti-social behavior and as a low density effect as seen from Fig.~\ref{fig-density}.

\rt{The study presented here should also be contrasted with previous studies on active particles with density dependent motility~\cite{cates2008,farrell2012pattern,cates2015motility,bauerle2018self,rein2016collective}. In fact, a class of models which exhibit MIPS are based on density dependent motility~\cite{solon2018generalized}. In such models, the motility depends on the particle density which is averaged over some region~\cite{solon2018generalized}. In the model presented in this article, we measure a very local and single-particle property, namely the existence or non-existence of a neighbor. The self propulsion speed of a particle does not change even if the number of neighbors changes, as long as there exists at least one neighbor.  
}

\rt{
Remarkably, it was shown experimentally and via numerical simulations that clustering emerges in a system of quorum sensing particles whose motility switches between an active and a passive state depending on the local concentration of signaling molecules~\cite{bauerle2018self}. In that case, a particle switches to a passive state whenever the local concentration of signaling molecules exceeds a threshold value. This is similar to the social case considered in this article. Our results suggest that it would be interesting to investigate whether clustering could emerge in the system considered in Ref.~\cite{bauerle2018self} also under anti-social conditions, i.e. if a particle chooses to switch to an active state when the concentration exceeds some threshold. 
}

Ref.~\cite{bechinger2019} showed that clustering can emerge based on perception of the environment in terms of distance to neighbors  without any active reorientation. The study presented here differs from that in Ref.~\cite{bechinger2019} in several ways. In the latter, the authors consider switching between active and passive states, whereas in our case the particles switch between two active states. In Ref.~\cite{bechinger2019} the choice of the mobility state depends on a perception function being greater than a threshold value. This function depends on the distance to all the other particles within the perception range and decays with distance, while the perception range is in general anisotropic. In this study the interaction range is fixed, and isotropic and the choice of mobility state depends only on the detection of a single neighbor within the interaction range, with no regard to the number of neighbors detected. We show that switching between two active states depending on the minimal detection of whether or not a neighbor is present within a fixed, isotropic interaction range is sufficient to result in cluster formation.  We consider both social and anti-social interactions, whereas in Ref.~\cite{bechinger2019}  the active state always encourages group formation, i.e. socially inclined behavior. Thus, the study presented here complements that in Ref.~\cite{bechinger2019}.

In our model, the interactions between the particles are non-reciprocal in nature. For example, consider three particles within an interaction range $R$, and therefore moving with a velocity of magnitude $V_i$. If one of the particles moves out of the interaction range, it will change its velocity to $V_o$, whereas the other two particles will continue to move with velocity $V_i$ as long as they remain within the interaction range of each other. Thus, the first particle ``feels'' its interaction with the other two, whereas the motion of these other two particles is not affected by the fact that the first particle moved away from them. Similarly, if two particles are already within a distance $R$ one from the other, and a third particle enters the interaction range of one of them, this third particle changes its mode of motion due to this interaction, while the first two particles are not affected. Such non-reciprocal interactions play a central role in synthetic active matter~\cite{brandenbourger2019, lea2020, fruchart2021}. The role of non-reciprocity in our model is a matter of future investigation which could lead to better understanding of the types of individual interactions that can give rise to different collective phenomena.

This study lays the ground for the realization of self-assembly and anti-clustering of robotic swimmers due to virtual interactions. By implementing simple sensors such as photo-diodes, one can introduce the ability of individual swimmers to sense whether there are other swimmers in their vicinity. Alternatively, one could realize the density function presented in this study by a minimal sensory platform. While our work stems from a statistical mechanics perspective and thus focuses on multiple swimmers, we observe clustering and anti-clustering already with numbers of particles as small as two. This can allow to initially test the interactions between a pair of robots and later extend it to larger groups. To overcome the statistical gap, one can examine the interactions over an extended duration of time, or to repeat the physical experiments multiple times. An existing robotic platform, capable of autonomous regulation of vertical buoyancy~\cite{kobo2022}, can be adapted to introduce propulsion in the horizontal plane. This will initially enable to study cooperative phenomena due to \rt{quorum sensing based} interactions in 2D, either at the water-air interface for positively buoyant robotic swimmers or in a specific plane underwater for neutrally buoyant swimmers. Later, this platform could be used to investigate 3D motion and to compare the physical differences between the interactions in two and three dimensions. In addition, it will enable robots to toggle between different behaviors in response to external triggers, as hypothesized for the behavior of fish~\cite{herbert2011}. While in this model inertial forces are not taken into account, in real robotic swimmers, one would need to address the flow regime, specifically the Reynolds number, in which the swimmers operate. This, however, opens new horizons to study collective behavior across different scales of underwater locomotion. In practice, this could be adjusted by the motors velocity or viscosity of the surrounding liquid.

\begin{acknowledgments}
We thank Guy Gabrieli, Michael A. Lomholt, Priyanka, and Yael Roichman  for insightful discussions. This research was supported by the Ministry of Science \& Technology, Israel, grant number 3-17384. ST acknowledges support in the form of a Sackler postdoctoral fellowship and funding from the Pikovsky-Valazzi matching scholarship, Tel Aviv University. 
\end{acknowledgments}

\appendix

\section{Supplementary Figures}
\label{sec:appendix}

\begin{figure*}[]
\includegraphics[scale=0.4]{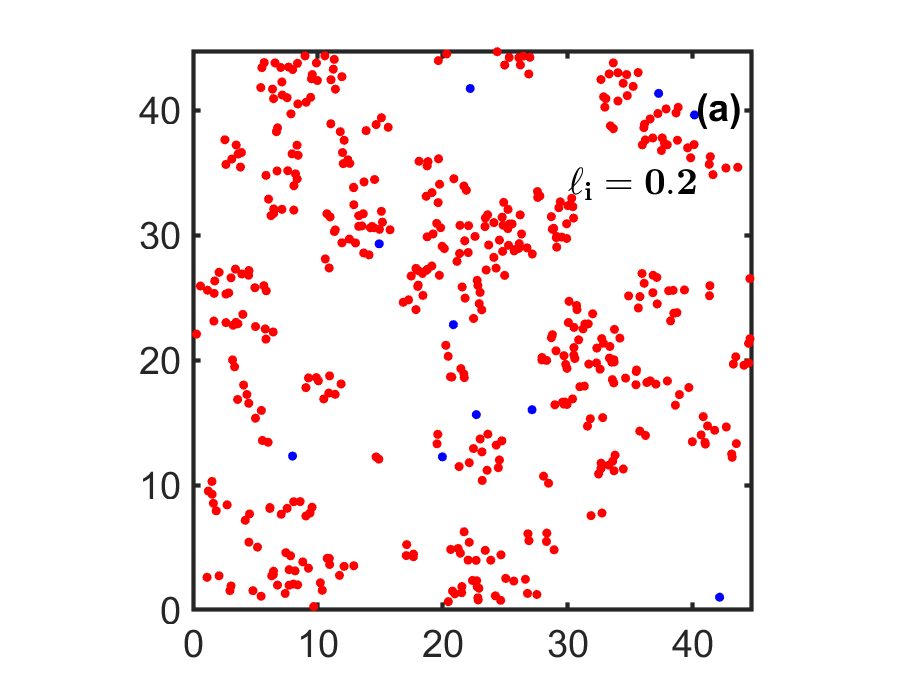}
\hspace{-1.2cm}
\includegraphics[scale=0.4]{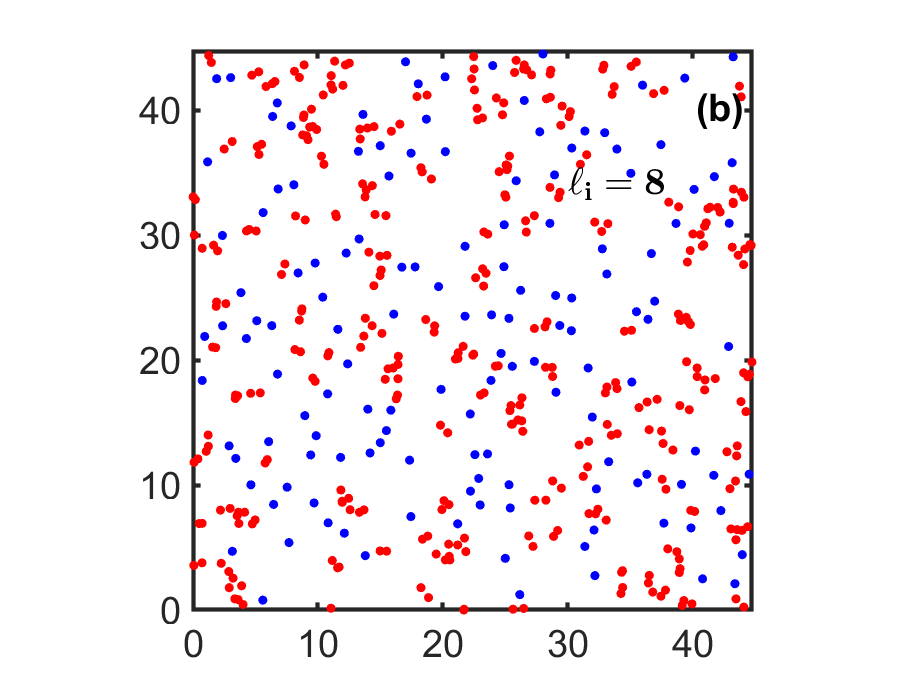}
\hspace{-1.2cm}
\includegraphics[scale=0.4]{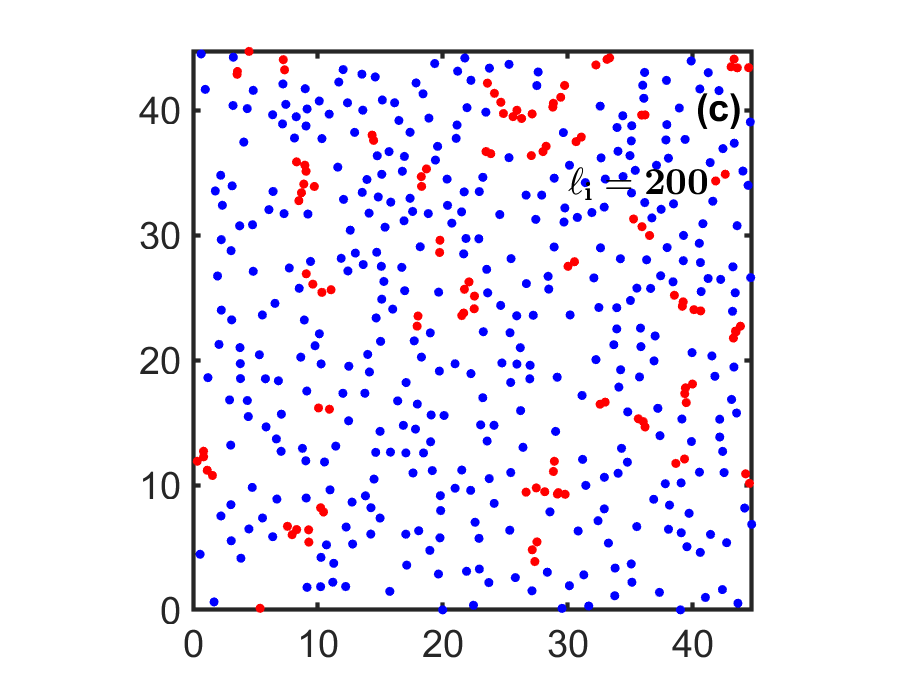}
\caption{Figure analogous to Fig.~\ref{fig-spos}, but with $N=500$. Representative realization of the positions of the particles, normalized by the interaction radius, in the 2D steady state with $\ell_o=20$ and $\ell_N=2$, and with (a) $\ell_i=0.2$, (b) $\ell_i=8$ and (c) $\ell_i=200$. (a) and (b) correspond to social behavior i.e. $\ell_i<\ell_o$, while (c) corresponds to anti-social behavior, i.e. $\ell_i>\ell_o$.}
\label{fig-spos500}
\end{figure*}

\begin{figure*}[]
\includegraphics[scale=0.44]{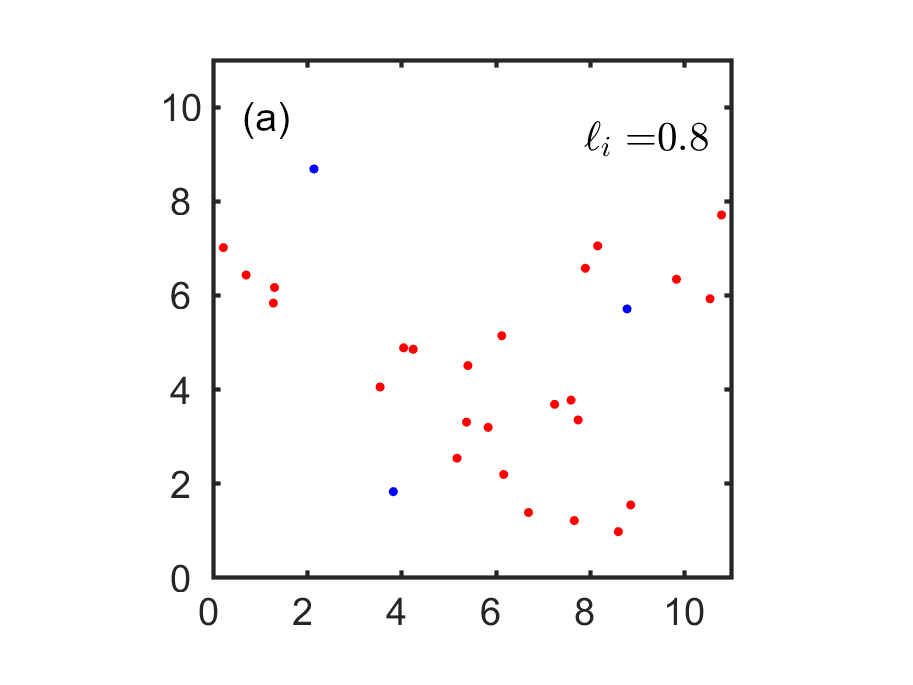}
\hspace{-1.5cm}
\includegraphics[scale=0.44]{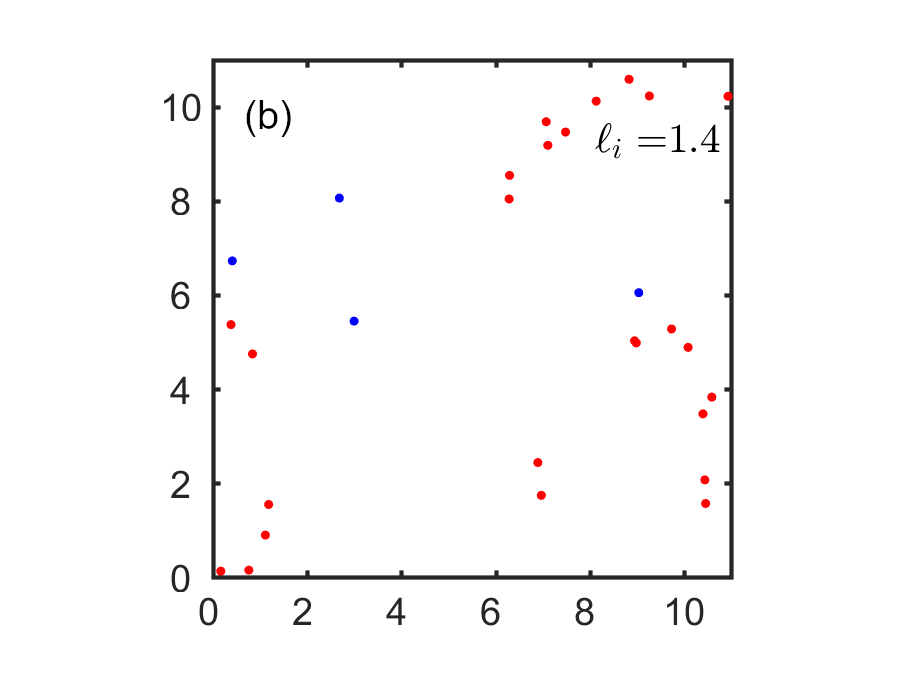}
\hspace{-1.5cm}
\includegraphics[scale=0.44]{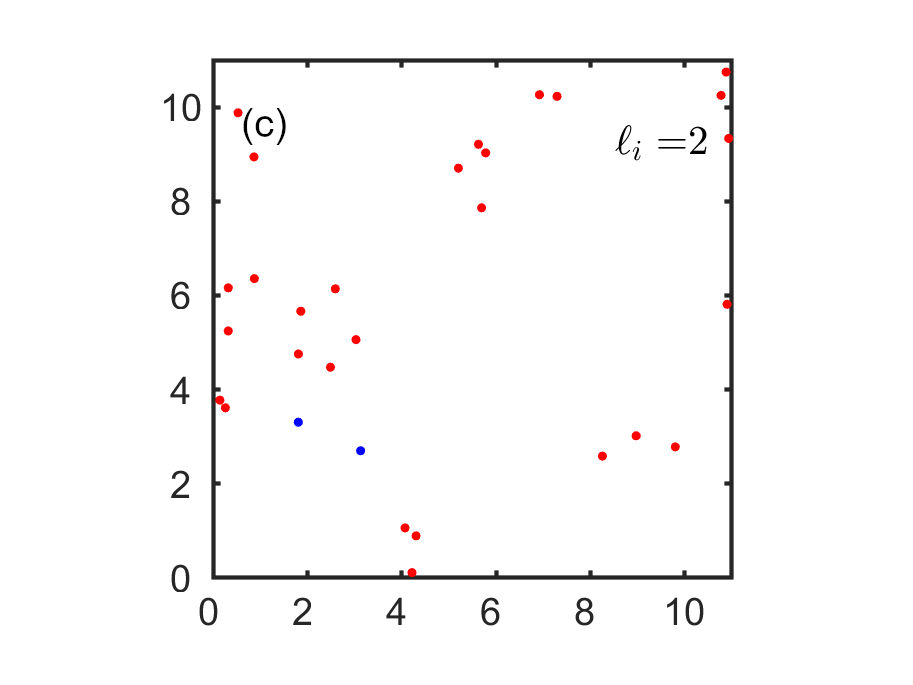}\\
\includegraphics[scale=0.44]{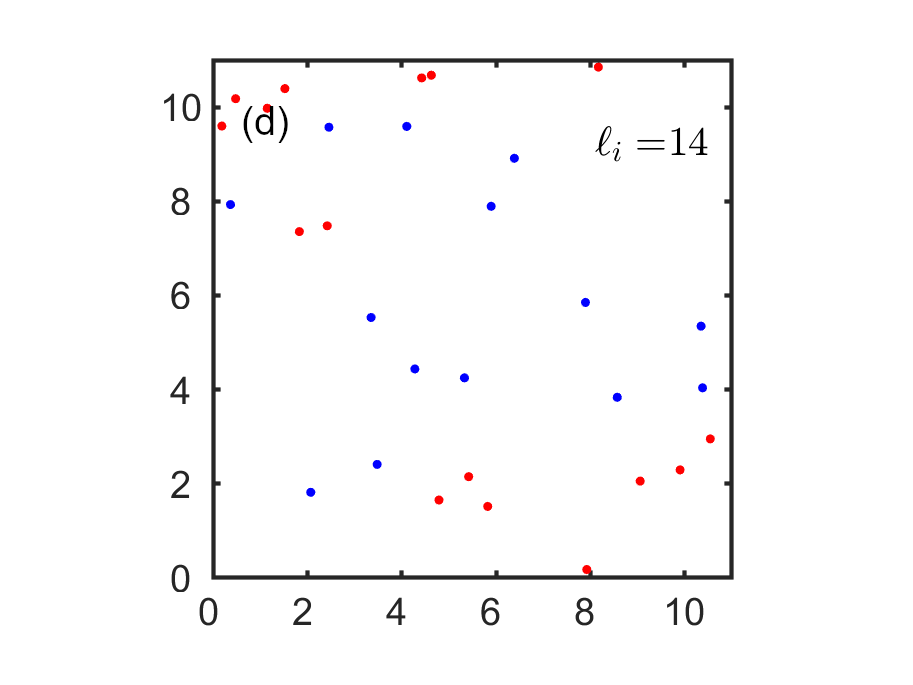}
\hspace{-1.5cm}
\includegraphics[scale=0.44]{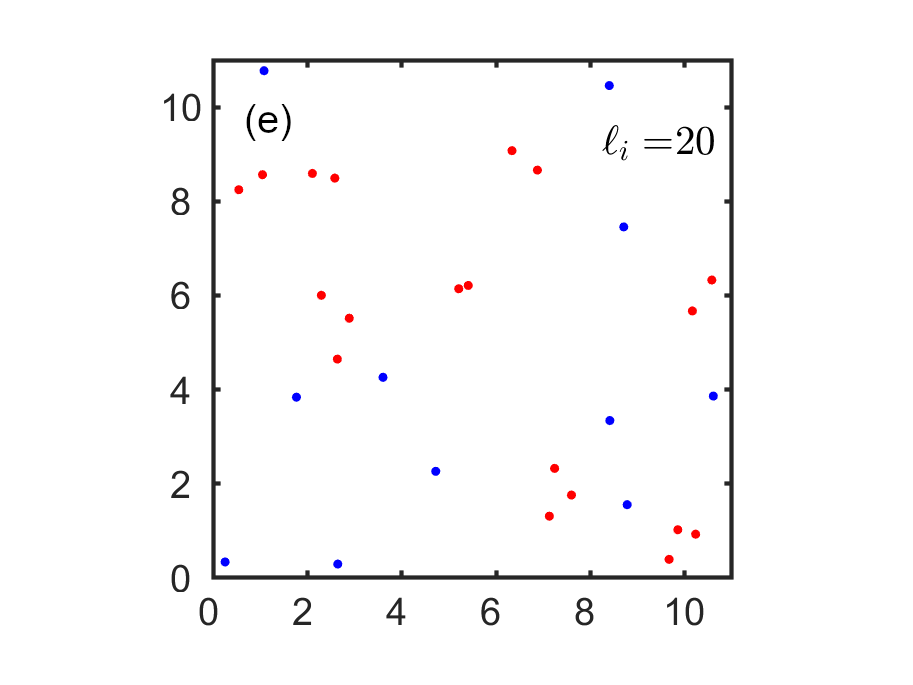}
\caption{Steady state positions of the particles in 2D steady state for} social behavior, $\ell_i < \ell_o$, all with $\ell_o=20$, $\ell_N=2$ and $N=30$.
\label{fig-social-pos}
\end{figure*}

\begin{figure*}[]
\includegraphics[scale=0.44]{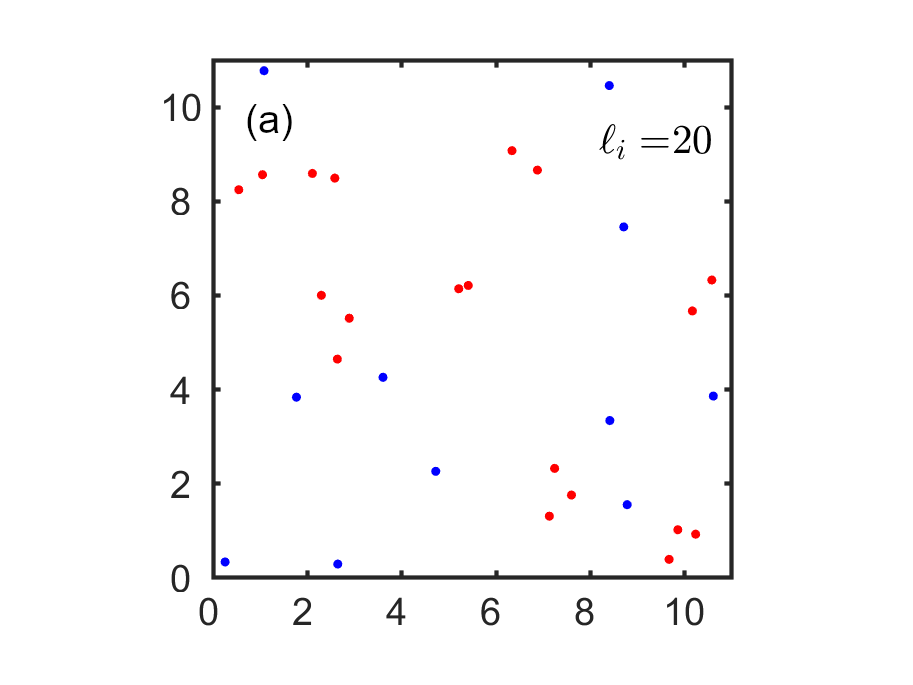}
\hspace{-1.5cm}
\includegraphics[scale=0.44]{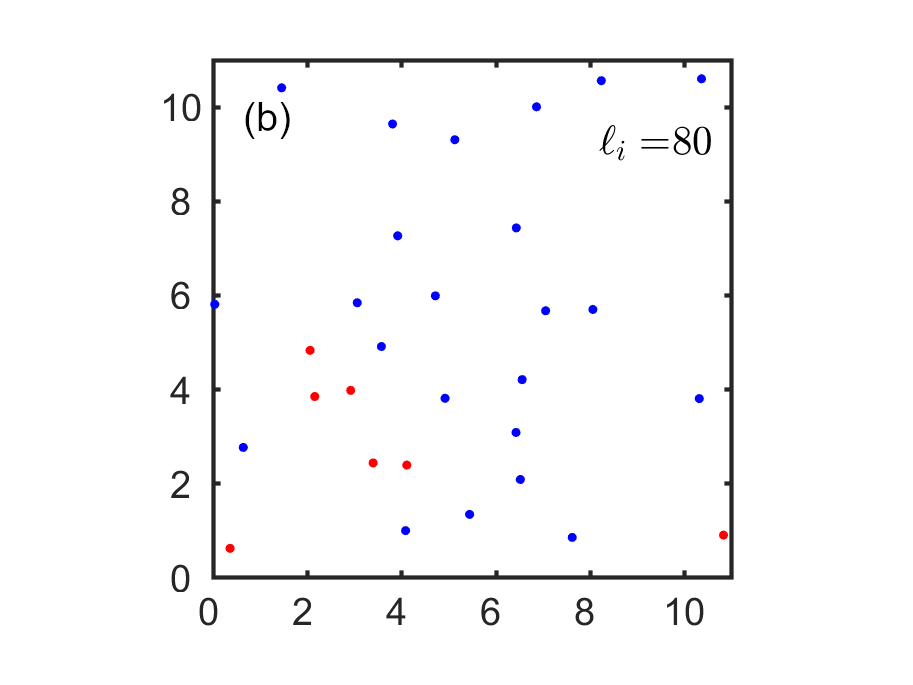}
\hspace{-1.5cm}
\includegraphics[scale=0.44]{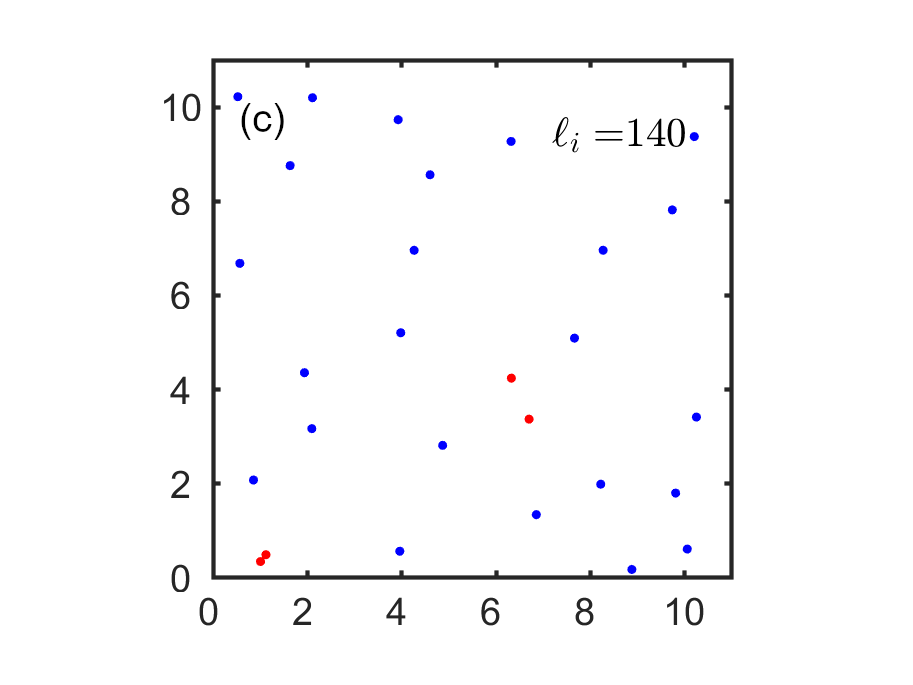}\\
\includegraphics[scale=0.44]{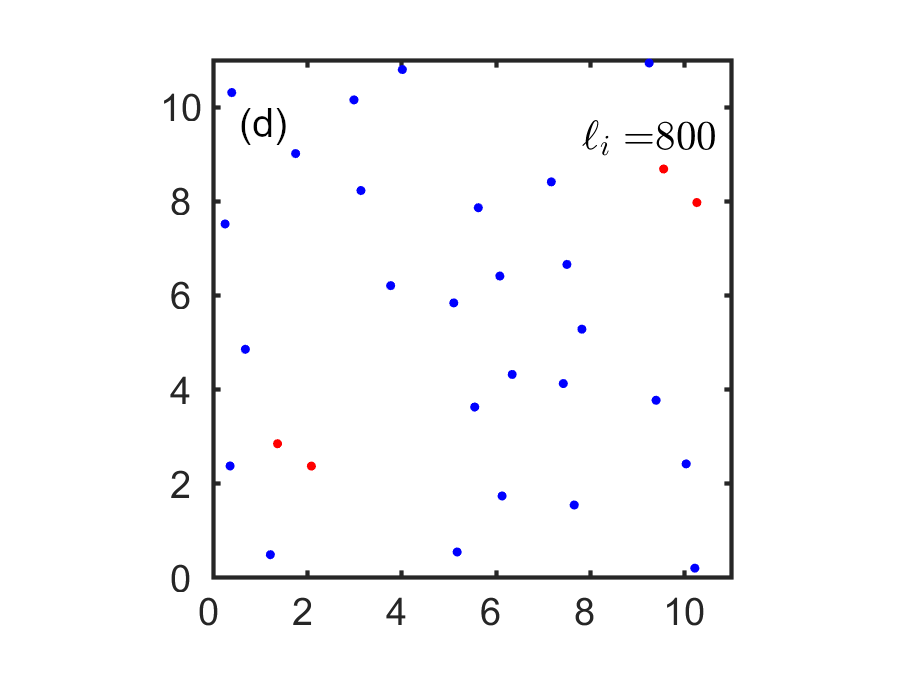}
\hspace{-1.5cm}
\includegraphics[scale=0.44]{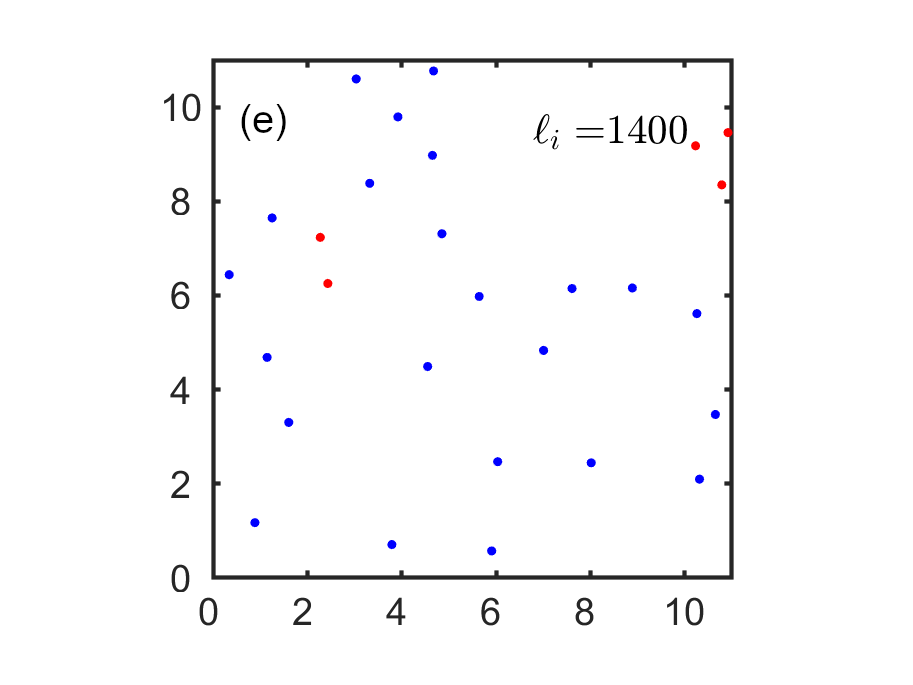}
\hspace{-1.5cm}
\includegraphics[scale=0.44]{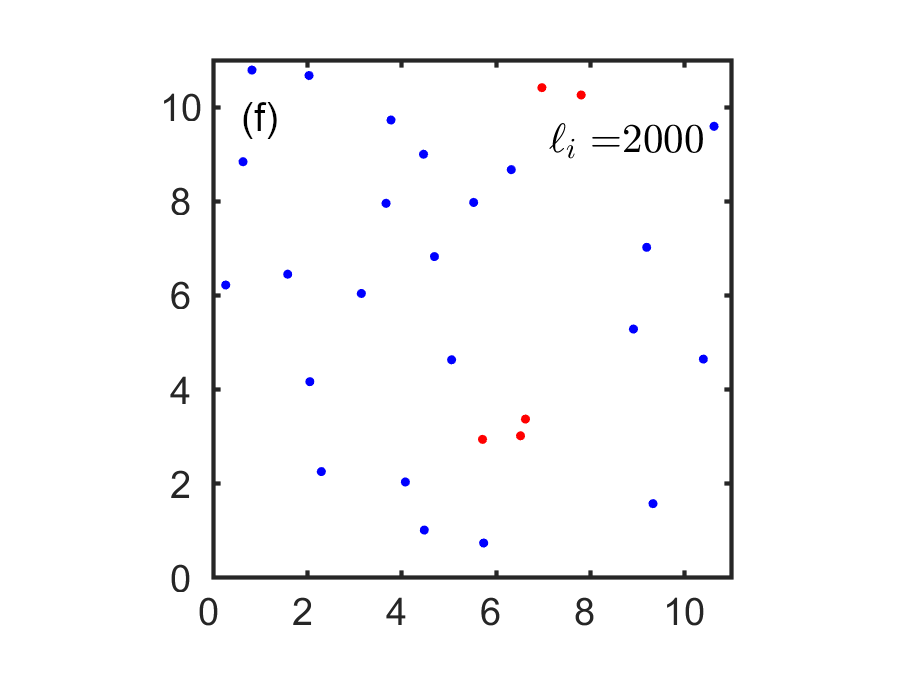}
\caption{Steady state positions of the particles in 2D steady state for anti-social behavior, $\ell_i > \ell_o$, all with $\ell_o=20$, $\ell_N=2$ and $N=30$.} 
\label{fig-asocial-pos}
\end{figure*}

\begin{figure*}[]
\includegraphics[width=0.8\columnwidth]{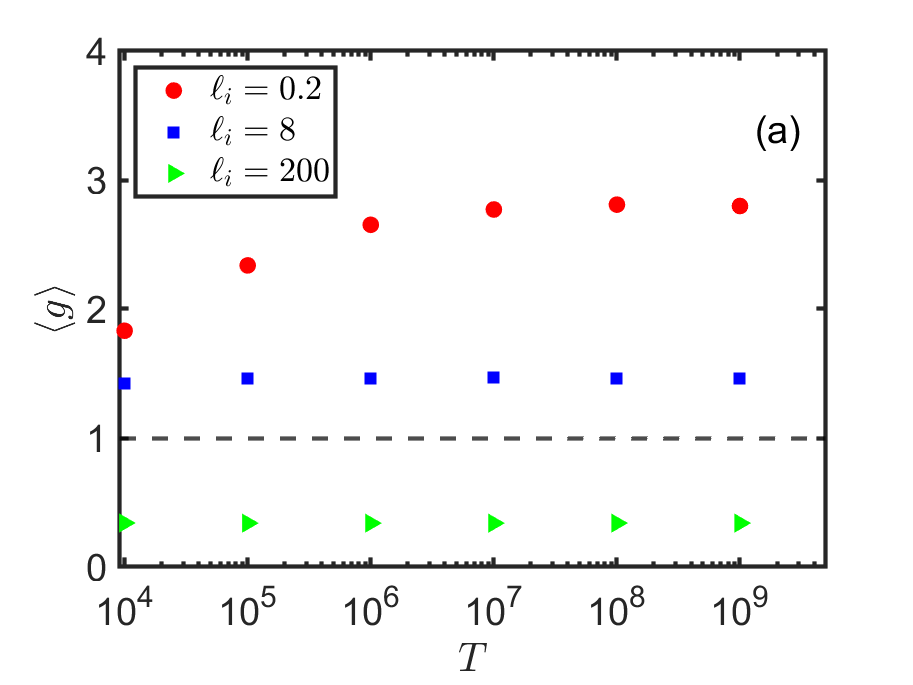}
\includegraphics[width=0.8\columnwidth]{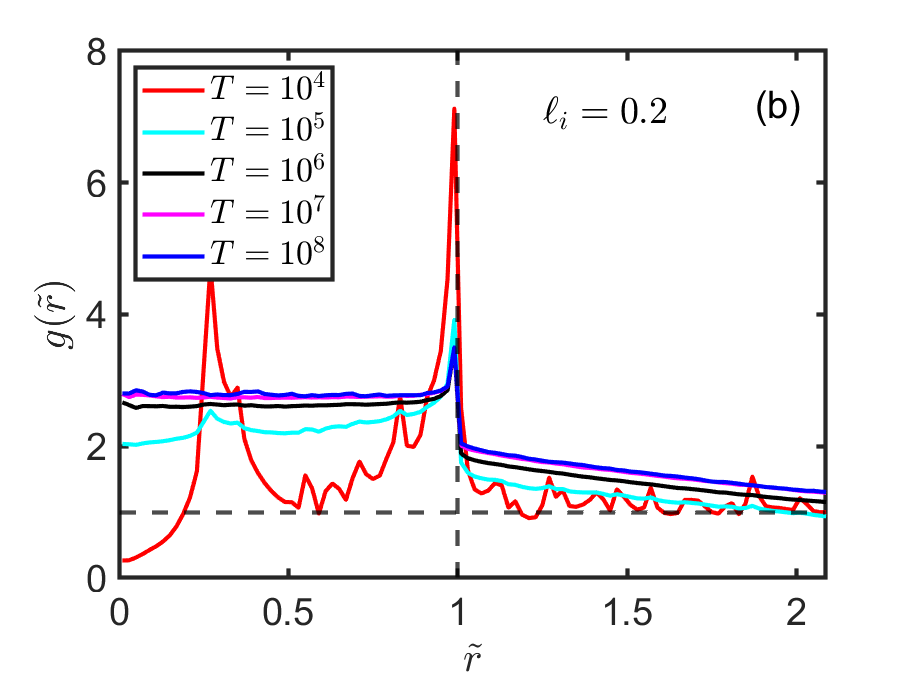}\\
\includegraphics[width=0.8\columnwidth]{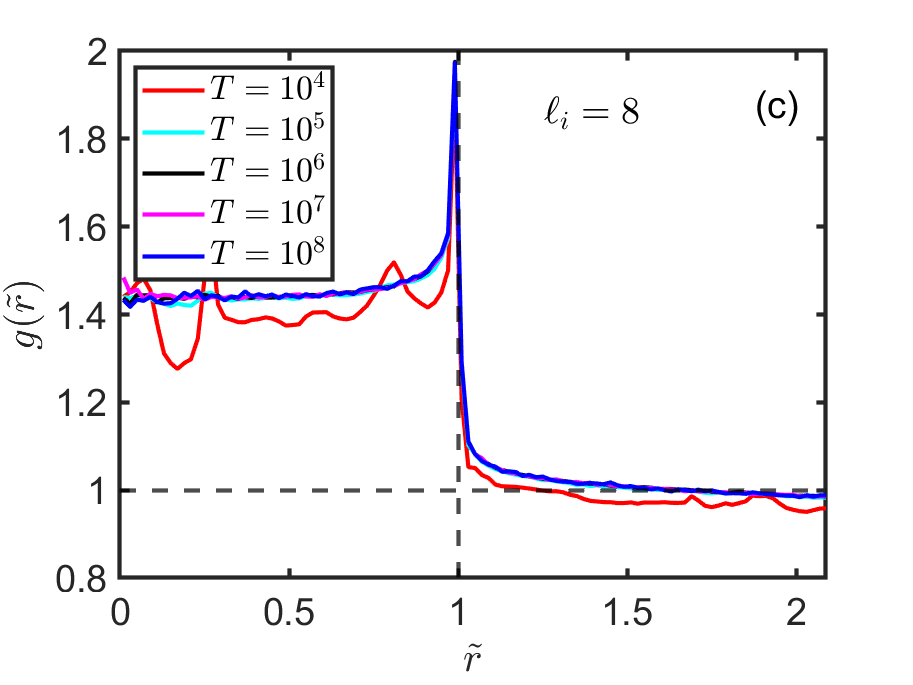}
\includegraphics[width=0.8\columnwidth]{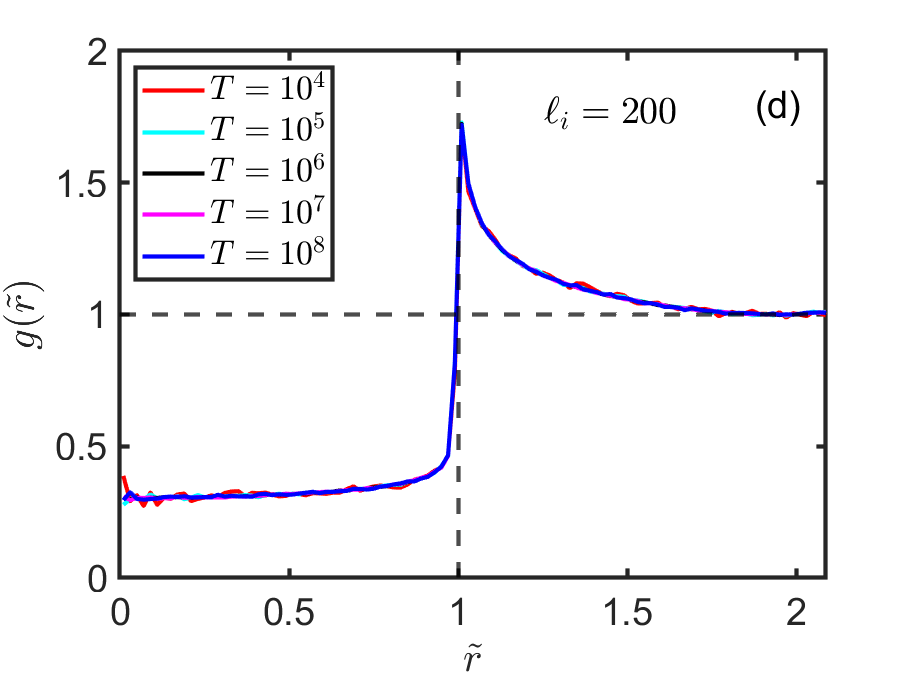}
\caption{Convergence of $g(r)$ to the steady state in 2D simulations with $N=30$ particles. (a) Average $\langle g \rangle$ within the interaction radius as a function of total run time $T$ for different $\ell_i$ as shown in the legend. Here, $\ell_N=2$ and $\ell_o=20$. The radial distribution function $g(\tilde{r})$ vs. the normalized distance between particles $\tilde{r}=r/R$, for different run times $T$ with (b) $\ell_i=0.2$, (c) $\ell_i=8$, and (d) $\ell_i=200$. Already with $T=10^5$, $g(\tilde{r})$ converges to the steady state, except for the case $\ell_i=0.2$ for which a run time of at least $T=10^7$ is needed to reach the steady state. The dashed line corresponding to $\tilde{r}=1$ shows the interaction radius. Note that each curve in (b), (c) and (d) is averaged over an ensemble of realizations starting with the same initial condition for the positions and the orientations. Thus, effects of limited statistics are eliminated, and the plots demonstrate the convergence to the steady state as a function of the total run time.}
\label{fig-gr-steady}
\end{figure*}

\begin{figure*}[]
\includegraphics[width=0.8\columnwidth]{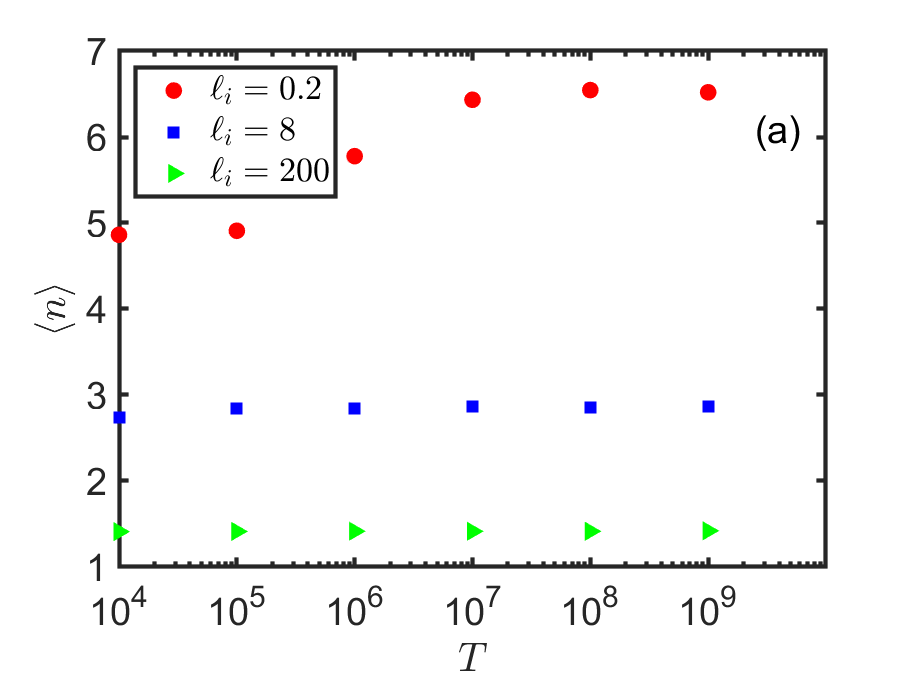}
\includegraphics[width=0.8\columnwidth]{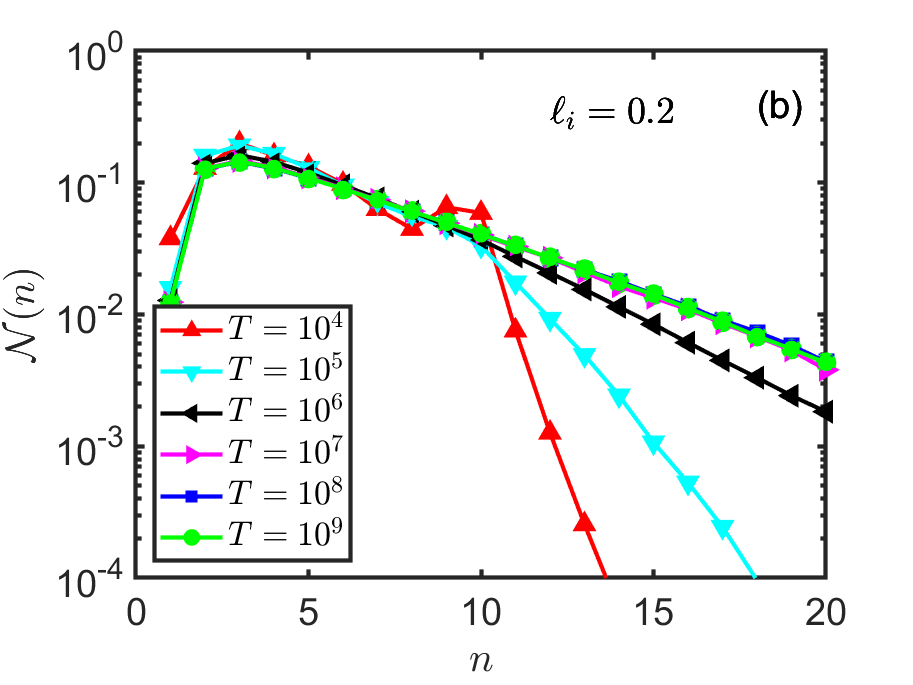}\\
\includegraphics[width=0.8\columnwidth]{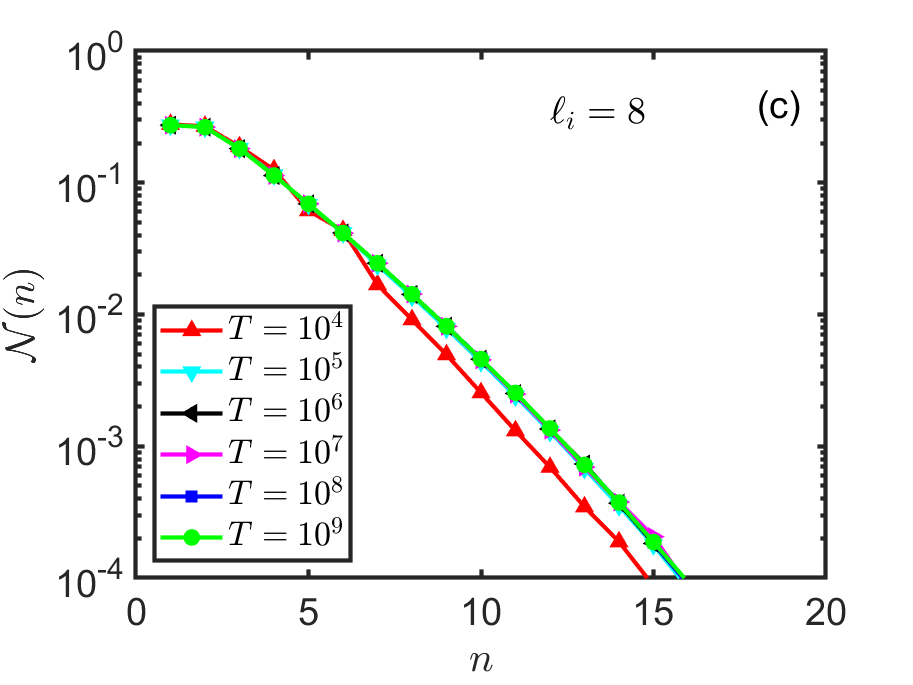}
\includegraphics[width=0.8\columnwidth]{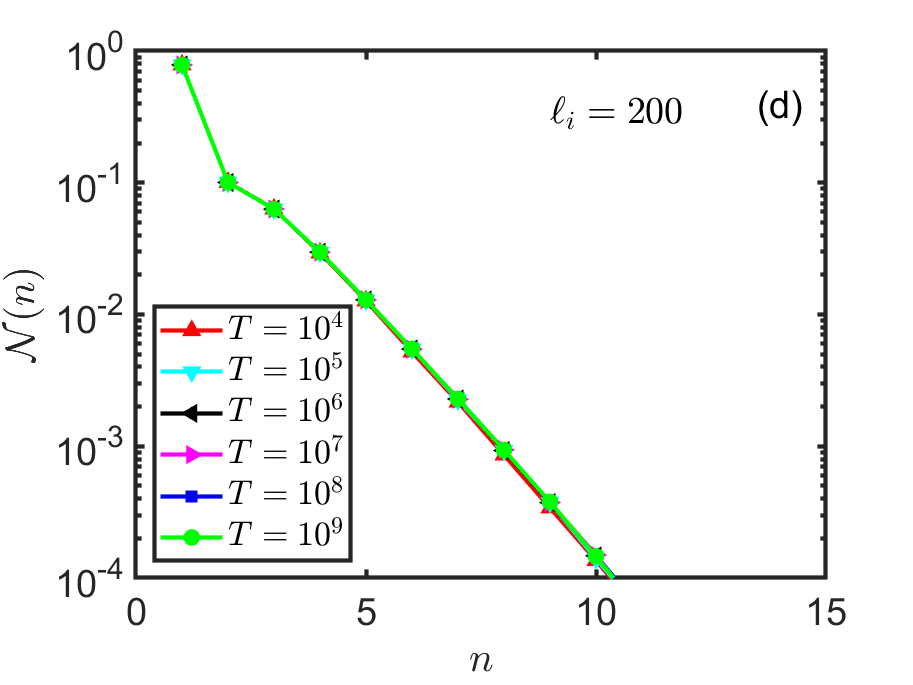}
\caption{Convergence of cluster size distribution to the steady state in 2D simulations with $N=30$ particles. (a) Average $\langle n \rangle$ as a function of total run time $T$ for different $\ell_i$ as shown in the legend. Here, $\ell_N=2$ and $\ell_o=20$. The cluster size distribution $\mathcal{N}(n)$ vs. the cluster size $n$, for different run times $T$ with (b) $\ell_i=0.2$, (c) $\ell_i=8$, and (d) $\ell_i=200$. Already with $T=10^4$, $\mathcal{N}(n)$ converges to the steady state, except for the case $\ell_i=0.2$ for which we a run time of at least $T=10^7$ is needed to reach the steady state. Note that each curve in (b), (c) and (d) is averaged over an ensemble of realizations starting with the same initial condition for the positions and the orientations. Thus, effects of limited statistics are eliminated and the plots demonstrate the convergence to the steady state as a function of the total run time.}
\label{fig-clust-steady}
\end{figure*}

Figure~\ref{fig-spos500} shows the steady state positions for the parameter values same as Fig.~\ref{fig-spos}, but with $N=500$. Figure~\ref{fig-social-pos} shows the positions of the particles evolving according to social interaction rules, i.e. $\ell_i<\ell_o$, all with $\ell_o=20$, $\ell_N=2$ and $N=30$. Figure~\ref{fig-asocial-pos} shows the positions of the particles evolving according to anti-social interaction rules, i.e. $\ell_i>\ell_o$, all with $\ell_o=20$, $\ell_N=2$ and $N=30$. Figure~\ref{fig-gr-steady} shows the convergence of the radial distribution function $g(r)$ to the steady state. Figure~\ref{fig-clust-steady} shows the convergence of the cluster size distribution $\mathcal{N}(n)$ to the steady state. 

\newpage

\newpage
\clearpage

\bibliography{swimmers_nourl}
\bibliographystyle{apsrev}

\end{document}